\documentclass[aps,twocolumn,showpacs,amsmath,amssymb,pra,superscriptaddress,floatfix]{revtex4-1}

\usepackage{graphicx}% Include figure files
\usepackage{dcolumn}% Align table columns on decimal point
\usepackage{bm}% bold math
\usepackage{amssymb}
\usepackage{multirow}
\usepackage{braket}
\usepackage{leftidx}
\usepackage{color}

\newcommand{\bs}{\beta_{6}}
\newcommand{\aperp}{a_{\perp}}
\newcommand{\operp}{\omega_{\perp}}
\newcommand{\ps}[1]{\delta_{#1}}
\newcommand{\kmat}{$K$ matrix }

\newcommand{\ttc}[2]{T_{#2}^{(#1)}}
\newcommand{\tcir}[2]{T_{\textrm{CIR},#1}^{(#2)}}
\newcommand{\tr}[1]{\operatorname{Tr}[#1]}

\newcommand{\transwf}[1]{\phi_{#1}(\rho)}
\newcommand{\abar}[1]{\bar{a}_{#1}}

\newcommand{\cs}{$\leftidx{^{133}}\rm Cs$ }

\newcommand{\lspwa}[1]{$#1$-SPWA}

\newcommand{\estar}[1]{\ep^{*}_{#1}}

\newcommand{\np}{n^{\prime}}
\newcommand{\nnp}{n\np}

\newcommand{\f}[1]{\mathbf{f}_{#1}}
\newcommand{\fo}[1]{\mathbf{f}_{#1}^{\mathbf{o}}}
\newcommand{\fc}[1]{\mathbf{f}_{#1}^{\mathbf{c}}}

\newcommand{\Fo}[1]{\mathbf{F}_{#1}^{\mathbf{o}}}
\newcommand{\Fc}[1]{\mathbf{F}_{#1}^{\mathbf{c}}}

\newcommand{\K}[1]{\mathbf{K}_{#1}^{1D}}
\newcommand{\Kp}[1]{\mathbf{K}_{#1}^{1D,phys}}
\newcommand{\KtD}{\mathbf{K}^{3D}}

\newcommand{\Umat}[1]{\mathfrak{U}_{#1}}
\newcommand{\cp}[2]{c_{#1}^{(#2)}}
\newcommand{\elp}{\ell^{\prime}}

\newcommand{\hurz}[2]{\zeta_{H}\big(#1,#2\bigr)}

\newcommand{\ep}{\epsilon}
\newcommand{\epc}{\epsilon_{c}}
\newcommand{\dep}{\Delta\epsilon}

\newcommand{\limU}[2]{\lim_{#1 \nearrow #2}}
\newcommand{\limD}[2]{\lim_{#1 \searrow #2}}
\newcommand{\bO}[1]{\mathcal{O}(#1)}

\begin{document}

\title{An analytical approach to atomic multichannel collisions in tight harmonic waveguides}

\author{Benjamin He\ss}
    \email{bhess@physnet.uni-hamburg.de}
    \affiliation{Zentrum f\"ur Optische Quantentechnologien, Universit\"at Hamburg, Luruper Chaussee 149, 22761 Hamburg, Germany}

\author{Panagiotis Giannakeas}
    \email{pgiannak@physnet.uni-hamburg.de}
%    \affiliation{Zentrum f\"ur Optische Quantentechnologien, Universit\"at Hamburg, Luruper Chaussee 149, 22761 Hamburg, Germany}
    \affiliation{Department of Physics and Astronomy, Purdue University, West Lafayette, Indiana 47907, USA}

\author{Peter Schmelcher}
    \email{pschmelc@physnet.uni-hamburg.de}
    \affiliation{Zentrum f\"ur Optische Quantentechnologien, Universit\"at Hamburg, Luruper Chaussee 149, 22761 Hamburg, Germany}
    \affiliation{The Hamburg Centre for Ultrafast Imaging, Universit\"at Hamburg, Luruper Chaussee 149, 22761 Hamburg, Germany} 

\date{\today}


\begin{abstract} 
We perform an analytical investigation in the framework of generalized $K$ matrix theory of the 
scattering problem in tight isotropic and harmonic waveguides allowing for several open scattering channels.
The scattering behavior is explored for identical bosons and fermions, as well as for distinguishable particles, the
main aspect being the confinement-induced resonances (CIR) which are attributed to different partial waves.
In particular we present the unitarity bounds which emerge when considering a quasi one dimensional system. Unitarity
bounds are also given for the transition coefficients, which show the limitations for efficient transversal
(de-)excitations by means of CIRs. We analyze the CIR for $d$-waves and find the intriguing phenomenon of a
strong transmission suppression in the presence of more than one open channel, which represents an interesting regime
to be applied in the corresponding many-particle systems. The corresponding channel threshold singularities are studied and it is
shown that these are solely determined by the symmetry class of the partial wave.
\end{abstract}

\maketitle
%%%%%%%%%%%%%%%%%%%%%%%%%%%%%%%%%%%%%%%

\section{Introduction}
\label{sec:introduction}
Trapping ultracold atomic vapors in tight waveguides and thus effectively reducing their dimensionality has become a key concept
in the contemporary study of ultracold atomic few- and many-body systems, as exotic low-dimensional quantum phases
\cite{girardeau1960,kinoshita2004,paredes2004} such as the Tonks-Girardeau gas in one dimension (1D) or the
Berezinsky-Kosterlitz-Thouless transition in two dimensions (2D) are available. Besides these intriguing
phenomena, the reduction of dimensionality also allows for a novel mechanism to control the scattering physics of
two-body interactions. This was first pointed out by Olshanii \cite{olshanii1998} who studied the influence of a tight
cylindrical confinement on scattering events. In particular, a resonance appears when the $s$-wave scattering length
becomes comparable with the length scale associated with the confining waveguide potential. This so called confinement-induced
resonance (CIR) as a result can be controlled by tuning the trap frequency, which recently also led to the first
experimental realization of a super Tonks-Girardeau gas \cite{haller2009,haller2009}. A similar prediction of a CIR for
spin-polarized fermions \cite{granger2004} was in the following also experimentally confirmed
\cite{guenter2005,moritz2005,froehlich2011}. Except these experiments discovering CIRs in 1D and 2D, a recent experiment
\cite{lamporesi2010} was carried out in mixed-dimensions.\par

However, the substantial theoretical effort prevail the experimental observations, while at the same time building up a
comprehensive understanding of the principles of CIR and suggesting a variety of systems where CIRs can emerge. These
efforts include works on different waveguide geometries
\cite{petrov2001,pricoupenko2006,idziaszek2006,peano2005,sala2012,peng2011,zhang2013quasi1dscatt}, resonant molecule
formation \cite{melezhik2009}, transparency induced by the confinement \cite{kim2006,hess2014}, CIRs in mixed dimensions
and multiple open channels \cite{saeidian2008,melezhik2011,nishida2010}, the coupling of various partial waves due to
the confinement \cite{giannakeas2012}, or atomic scattering with anisotropic interactions
\cite{sinha2007,giannakeas2013}.  Alongside with these studies we also want to mention recent investigations on ultracold quantum
gases on atom chips \cite{buecker2011,buecker2012,vanfrank2014} where excited transversal modes of the confinement are
utilized in order to prepare entangled atom clouds.\par 

The focus of the present study is the two-body multi-mode scattering behavior of atoms in the presence of an axially
symmetric and harmonic waveguide. In addition to the treatment of bosonic and spin-polarized fermionic collisions, we
also provide a theoretical description of the collisional properties of distinguishable particles, for which even and
odd partial waves contribute.  Similar to preceding studies \cite{giannakeas2012,hess2014}, the particles are allowed to
interact with higher partial waves. The constraint of colliding with energies below the excitation energy of the first
excited transversal state is however lifted in this work, thus allowing inelastic collisions, where particles can be
scattered into different channels. The latter are asymptotically defined by the transversal trap modes. Our approach is
based on the fully analytical and non-perturbative description in terms of $K$ matrices, whose usefulness has already
been demonstrated in a series of previous works \cite{granger2004,giannakeas2012,giannakeas2013,hess2014}.  Due to the
seminal results of Bo Gao \cite{gao1998solvdw,gao1998qdtvdw,gao2001angular} who investigated the free-space scattering
properties of neutral alkaline atoms possessing a van der Waals tail, analytical formulas for the (generalized)
scattering length were derived, which are used in the present setup to adequately describe the scattering event on the
interatomic scale.

The $K$ matrix approach provides a generalization of the works of Granger {\it et al}
\cite{granger2004} and Kim {\it et al} \cite{kim2005} incorporating however {\it all} the higher partial waves and
contributions from {\it all} the closed channels. Furthermore, going beyond the previous studies we derive the
connection of the {\it physical} $K$-matrix with all the relevant scattering observables obtaining thus the {\it full }
multi-component scattering wave function. Using this formalism we study the universal properties of ultracold
collisions in the $\ell$-wave single partial wave approximation (\lspwa{\ell}). Here, we find the existence of energies
above the corresponding channel threshold at which the collision effectively behaves as in free-space. This energy is
found to be independent of the number of open channels for $s$- and $p$-wave interactions. Furthermore, we investigate
the unitarity bounds for quasi-1D collisions. This quantitatively explains the transmission at a CIR in higher
transversal modes and explains the confinement induced unitarity bound, from which we also derive the unitarity bounds
for inelastic collisions in the waveguide. These bounds may be useful when investigating the possibilities and
limitations of populating higher transversal modes by means of a CIR, as interest in coherent excitations in waveguides
increases \cite{buecker2011,buecker2012,vanfrank2014}.  Next, we investigate the intriguing possibility of a blockade in
the first excited transversal mode, found from the quasi 1D unitarity bound in the \lspwa{d}, with an adequate
interatomic potential and the coupling of partial waves taken into account, showing that an almost totally blockaded
transmission channel may exist even if there are other possible scattering channels available. In addition we discuss
the scattering of distinguishable particles in waveguides and show the qualitative difference of partial wave coupling
for distinguishable particles by introducing resonance and transparency coefficients \cite{hess2014}. We discuss the
occurring threshold singularities for collisions of indistinguishable particles and show the qualitatively different
behavior for bosons and fermions.\par

In detail, this paper is organized as follows. Section \ref{sec:model} gives a brief review of the considered waveguide
Hamiltonian as well as the applied techniques, namely the $K$ matrix formalism while employing the local frame
transformation. Thereafter, Sec. \ref{sec:scattering_observables} introduces and discusses the relevant scattering
observables for 1D multichannel collisions as well as provides their connection to the physical $K$ matrix for the case
of 1D geometries. Section \ref{sec:results_discussion} is devoted to the analysis of our results while our summary and
conclusions are given in Sec. \ref{sec:conclusions}. The Appendix provides among others some technical concepts used to
derive the physical $K$ matrix.

\section{Waveguide Hamiltonian and $K$-matrix approach}
\label{sec:model}
In the following we study the collisional behavior of two particles within a harmonic waveguide. Hereby
indistinguishable or distinguishable particles are considered. The harmonic nature of the confining potential does not
couple the center of mass (CM) coordinates with the relative ones and permits us to treat their motions separately. The
Hamiltonian accounting for the CM motion simply describes a CM excitation in a harmonic potential. This solution is well
known and thus of no further interest. The non-trivial part is the relative motion Hamiltonian, which reads
%--
\begin{equation}
H=\frac{-\hbar^2}{2\mu}\Delta+\frac{1}{2}\mu\omega_{\perp}^2\rho^2+V_{LJ}(r),
\label{eq:Hamiltonian}
\end{equation}
%--
where $r=\sqrt{z^2+\rho^2}$ is the interparticle distance, with $z$ and $\rho$ describing the longitudinal and
transversal degrees of freedom, respectively.  $\mu$ denotes the reduced mass of the colliding pair and $\omega_\perp$
is the confinement frequency.  Accordingly, the harmonic oscillator length scale is given by $a_{\perp}=\sqrt{\hbar /
 \mu \omega_{\perp}}$.  The term $V_{LJ}(r)=\frac{C_{10}}{r^{10}}-\frac{C_{6}}{r^{6}}$ is the Lennard-Jones 6-10
potential indicating the short-range, two-body interatomic interactions.  $C_{6}$ is the dispersion coefficient and it
defines the van der Waals length scale via the relation $\beta_{6}=(2\mu C_{6}/\hbar^2)^{1/4}$.
We regard $C_{10}$ as a parameter in order to tune the corresponding scattering lengths induced by the short-range
potential. Among others, the particular choice $V_{LJ}(r)$ is motivated by the fact that we attribute to the two-body
physics a realistic character avoiding the use of zero-range approximations. In addition, this particular type of
interactions are analytically solvable by means of the generalized effective range theory \cite{gao2009single}. However,
any other interatomic potential is also suited as long as the length scale $\beta_n$ associated with this potential is
small compared to the oscillator length, i.e. $\beta_n\ll a_{\perp}$.\par

As in previous works \cite{granger2004,giannakeas2012,giannakeas2013,hess2014} on the Hamiltonian given in Eq.
\eqref{eq:Hamiltonian}, the separation of length scales is assumed, i.e. $\bs\ll\aperp$. In short, this implies that the
Hamiltonian has three distinct regions where (a) different potential contributions dominate and (b) different
symmetries are obeyed by the corresponding Hamiltonian.\par

(i) Starting in the inner region, where $r\sim \bs$ holds, the interatomic potential dominates. The two particles thus
experience a free-space collision with total energy $E=\frac{\hbar^2 k^2}{2\mu}$.\par 

(ii) The effect of this collision on the wave function is best observed from the intermediate region $(\bs\ll r\ll
\aperp)$, where both potential contributions are negligible. Hence, we can monitor the outcome of the collisional event
in region (i) by a well defined phase shift $\ps{\ell}$ for each partial wave. Due to the invariance under rotations
$SO(3)$ we can arrange the full scattering information in a diagonal, energy dependent, K matrix $\KtD$.\par

(iii) As the asymptotic region $r\gg\aperp$ is concerned, only the transverse confining potential contributes. The wave
function is thus a direct product of a sine or cosine function in the $z$-direction and a 2D harmonic oscillator (HO)
mode for the $\rho$-direction, i.e. $\ket{\psi}=\sum_{n}c_n\ket{q_{n};n;m}$, where $q_{n}$ denotes the channel momentum
in $z$-direction, $n$ denotes the transversal oscillator mode and $m$ the magnetic quantum number. The total energy $E$
distributes over these two degrees of freedom according to the relation $E=\hbar\omega_{\perp}(2n+|m|+1)+\frac{\hbar^2
  q_{n}^2}{2\mu}$.\par

We note however, that the azimuthal quantum number $m$ asymptotically associated to the $SO(2)$ rotations, is a good
quantum number in both regimes and therefore is a fixed quantity throughout our analysis which is set to $m=0$ for what
follows and is therefore also omitted in the labeling of the states. Hereafter, we drop the typical assumption in most
of the existing literature that the total collision energy has to be sufficiently small such that only the energetically
lowest transversal state can be populated, i.e.  $\ket{\psi}\sim\ket{q_0;0}$ and thus we are going beyond previous
studies by allowing inelastic collisions involving several transverse modes. Already from this expansion which is only
invariant under $\mathcal{T}_z\otimes SO(2)$, where $\mathcal{T}_z$ denotes translation along the $z$-direction and
$SO(2)$ rotations around that axis, we conclude that the mapping from region (ii) to (iii) cannot be accomplished by a
unitary transformation. Thus, in order to transfer the scattering information between these two regions of different
symmetry, an appropriate way is given by the local frame transformation $U_{ln}$
\cite{harmin1982nonhydrostark,*harmin1982stark,*harmin1985,fano1981,greene1987}. However, as already discussed previously
\cite{granger2004,giannakeas2013} the application of this technique comes at the price of rendering the closed channel
$(\hbar \omega_\perp(2n+1)> E)$ components of the wavefunction unphysical. This drawback is due to the boundary
conditions of our scattering approach, namely the standing wave approach, which after the analytical continuation to the
closed channels turns the oscillating solutions into exponential diverging ones. To overcome this unphysical situation a
standard channel closing procedure, familiar from multichannel quantum defect theory has to be applied \cite{aymar1996},
leading to the physical K matrix, given by
%--
\begin{equation}
  \Kp{oo}=\K{oo}+i \K{oc}(1-i\K{cc})^{-1}\K{co},
  \label{eq:k_oo_1d_phys}
\end{equation}
%--
where in turn, $\K{}$ refers to the corresponding 1D $K$ matrix $\K{}=U^{T} \KtD U$
\cite{granger2004,giannakeas2012,giannakeas2013,zhang2013quasi1dscatt}.  In addition, $\K{oo},\K{cc}$ denote the
open-open and closed-closed channel transitions, respectively. Accordingly, $\K{oc}$ and $\K{co}$ denote the open-closed
channel transitions and vice versa. From Eq. \eqref{eq:k_oo_1d_phys} the resonant processes are given as poles of the
physical $\K{}$ matrix. Therefore, the roots of $\det(\openone-i\K{cc})$ correspond to the positions of the closed
channel bound states lying in the continuum of the open channels. The resonant structure thus fulfills a Fano-Feshbach
scenario \cite{chin2010}. Performing the above calculations (cf. Appendix \ref{app:kphys}) for a given $3D$ $K$ matrix
$K^{3D}=\operatorname{diag}(\tan\delta_{\ell},\tan\delta_{\elp})$ yields a physical $K$ matrix, given by
%---
\begin{align}
\Kp{oo}&=\frac{1}{\det(\openone-i
\KtD\Umat{})}\times\Bigl\{\Delta_\ell\Fo{\ell\ell}+\Delta_{\ell^{\prime}}\Fo{\ell^{\prime}\ell^{\prime}}-\nonumber\\
&-i\Delta_{\ell}\Delta_{\ell^{\prime}}\bigl(\Umat{\ell^{\prime}\ell^{\prime}}\Fo{\ell\ell}+\Umat{\ell\ell}\Fo{\ell^{\prime}\ell^{\prime}}-\Umat{\ell\ell^{\prime}}(\Fo{\ell\ell^{\prime}}+\Fo{\ell^{\prime}\ell})\bigr)\Bigr\},
\label{eq:kphys}
\end{align}
%---
where $\Delta_{\ell}(E)=\tan\delta_{\ell}(E)$ contains the energy dependent $\ell$-th phase shift and the matrices
$\Fo{\ell\elp}$ are given by $(\Fo{\ell\elp})_{\nnp}=U_{\ell n}U_{\elp \np}$, with $0\le n,\np\le n_o-1$ and $n_o$
denoting the number of open channels. At this point we also introduce the generalized, energy dependent, scaled
scattering length defined for all partial waves $\ell$, by
$\abar{\ell}(E)^{2\ell+1}=(a_{\ell}(E)/\aperp)^{2\ell+1}=-\frac{\Delta_{\ell}(E)}{(\aperp k)^{2\ell+1}}$.  As the local
frame transformation $U_{\ell n}$ also depends on the energy, this also holds for the matrices
$\Fo{\ell\elp}=\Fo{\ell\elp}(E)$. Here, we also note that one should carefully distinguish the number of open channels
$n_o$ from the actual quantum numbers of the corresponding transverse modes $n$, e.g. consider the single mode regime
where only the population of the lowest transversal mode $\ket{q_0;0}$ is allowed, we have $n_o=1$.\par

The energy dependent $K$ matrix $\Kp{oo}$, given in Eq. \eqref{eq:kphys} appropriately describes the scattering in a
tight harmonic waveguide with several open transverse modes. The coupling of two arbitrary partial waves $\ell$ and
$\elp$ belonging to the same symmetry class, i.e. $\ell-\elp=0\operatorname{mod}2$, is properly taken into account.
Before proceeding let us recall that the effect of the closed channels on the scattering phase shift can conveniently be
expressed as
%---
\begin{align}
  \Umat{\ell\elp}(\ep)&=\sum_{p=0}^{\ell+\elp}\frac{\cp{p}{\ell,\elp}}{2(\ep+\frac{1}{2})^\frac{p+1}{2}}\hurz{-\frac{p-1}{2}}{n_{o}-\ep},
\label{eq:closed_channel_umat}
\\
\cp{p}{\ell,\elp} &=
(-1)^{\frac{\ell+\elp}{2}}\sqrt{(2\ell+1)(2\elp+1)}\nonumber\\
&\times \sum_{\nu=\max\{p,|\ell-\elp|\}}^{\ell+\elp}\Gamma(\ell,\elp,\nu,p)\quad,
\label{eq:clossel_channel_expansion_coefficients_cll}
\\
\Gamma(\ell,\elp,\nu,p)&=i^{p-1}2^{\nu-1}(2\nu+1)\nonumber\\
&\times\begin{pmatrix}v\\p\end{pmatrix}\begin{pmatrix}\frac{\nu+p-1}{2}\\\nu\end{pmatrix}\begin{pmatrix}\ell&\elp&\nu\\0&0&0\end{pmatrix}
\label{eq:clossel_channel_gamma}
\end{align}
%---
where $\zeta_{H}(s,a)$ denotes the Hurwitz zeta function and $\Gamma(\cdot)$ are some combinatorial constants containing
the Wigner $3j$-symbols. Equation \eqref{eq:closed_channel_umat} is discussed in more detail in \cite{hess2014}.  The
dimensionless, channel-normalized energy $\ep$ is defined by the relation $E=2\hbar\operp(\ep+\frac{1}{2})$, which is
chosen such that $n\le\ep\le n+1$ is between the threshold of th $n$-th and the $(n+1)$-th channel. We note, that
$n_{o}=\lfloor\ep\rfloor+1$ is the number of open channels, where $\lfloor\ep\rfloor$ denotes the largest integer
smaller than $\ep$. The representation of Eq. \eqref{eq:closed_channel_umat}, which differs from the one introduced in
\cite{hess2014}, is in particular useful when considering the threshold singularities below. One further remark is in
order, which refers to the second argument of the Hurwitz zeta function. This argument is given by
$\epc=n_{o}-\ep=1-\dep$, where $\dep=\ep-\lfloor\ep\rfloor$ denotes the fraction of the total collision energy above the
threshold of the last open channel rendering $\zeta_{H}$ periodic with a saw-tooth like behavior. \par

Throughout the following analysis the energy ranges up to $\ep=4$, passing several channel thresholds of the transverse
confinement. We note that all these energies lie close to the threshold of the
interatomic potential and thus the analytic solutions \cite{gao2009single} for our interatomic potential are
applicable.\par

\section{Scattering observables}
\label{sec:scattering_observables}
In the asymptotics of the scattering process the transversal and longitudinal degrees of freedom are decoupled and
the quantum number of the transversal $2D$ HO modes can be used to define the asymptotic scattering channel.
%---
\begin{equation}
\psi_n(r)=e^{iq_{n}z}\transwf{n}+\sum_{n^{\prime}=0}^{n_o}f_{nn^{\prime}}^{\pm}e^{iq_{n^\prime}|z|}\transwf{n^{\prime}},
\label{eq:1d_scattering_wavefunction}
\end{equation}
%---
describing an incoming wave in channel $n$ which is then (in)-elastically scattered into all open channels.  Here, the
scattering amplitude $f_{nn^{\prime}}^{\pm}$ in forward $(+)$, respectively backward $(-)$ direction reads
%---
\begin{equation}
f_{nn^{\prime}}^{\pm}=f_{nn^{\prime}}^e+\operatorname{sgn}(z)f_{nn^{\prime}}^o,
\label{eq:scattering_forward_backward}
\end{equation}
%---
whereas in turn $f^{e}$ and $f^{o}$ refer to the respective scattering amplitudes for even and odd exchange symmetry
and, $\operatorname{sgn}(z)=z/|z|$ denotes the sign function. By the conservation of flux, the forward $f^+$ and
backward $f^-$ scattering amplitude contain the same information about a scattering
event, we concentrate our analysis on $f^{+}$, for which the transmission and
reflection coefficients $T_{\nnp}$ and $R_{\nnp}$, respectively, from channel $n$ to $\np$ take the following form
%---
\begin{align}
  T_{\nnp}^{(\ell,\elp)} &= |\delta_{\nnp}+f^{\ell}_{\nnp}+f^{\elp}_{\nnp}|^2
\label{eq:transmission_fplus}
\\
R_{\nnp}^{(\ell,\elp)} &= |f^{\ell}_{\nnp}+f^{\elp}_{\nnp}|^2,
\label{eq:reflection_fplus}
\end{align}
%---
where $\ell$ and $\elp$ refer to even and odd partial waves, respectively.  The transition probability
$W_{\nnp}^{(\ell,\elp)}$ characterizing the transversal excitation and de-excitation processes from channel $n$ into a
specific channel $\np$ are given by the sum of the corresponding transmission and reflection coefficients
%---
\begin{equation}
W_{\nnp}^{(\ell,\elp)}=T_{\nnp}^{(\ell,\elp)}+R_{\nnp}^{(\ell,\elp)}
\label{eq:transition_nnp}
\end{equation}
%---
If the constituents of the scattering event both belong to the same symmetry class, i.e. both are either bosons or
fermions, we obtain a special case of Eq. \eqref{eq:transmission_fplus}, given by
$T_{\nnp}^{(\ell)}=|\delta_{nn^{\prime}}+f_{\nnp}^{(\ell)}|^2$, where the scattering amplitudes $f_{\nnp}^{(\ell)}$ are
connected to the physical $K$ matrix, via
%---
\begin{equation}
  \boldsymbol f^{(\ell)}=i\;\Kp{oo,\ell} \left[\openone-i\;\Kp{oo,\ell}\right]^{-1},
\label{eq:connection_f_kphys}
\end{equation}
%---
see also \cite{hess2014}. This relation in particular allows for a extension of previous studies on the CIRs in harmonic
waveguides to distinguishable particles, as it is in detail discussed in Sec.  \ref{sub:distinguishable}.\par

\section{Results and Discussion}
\label{sec:results_discussion}

\subsection{Universal Properties}
\label{sub:universalities}
We start our discussion of the universal properties for the (in)-elastic scattering in waveguides by considering the
$\ell$-wave single partial wave approximation (\lspwa{\ell}). Since the main focus is the study of CIRs which typically
occur in the vicinity of a free space resonance, the \lspwa{\ell} can safely be assumed to be accurate in the
description of the scattering process. Recall that for the particular choice of a Lennard-Jones type 6-10 interatomic
potential the background scattering length from higher lying partial waves ($\ell>1$) is negligible.
The advantage of the \lspwa{\ell} is given by the fact, that (i) analytical results can be derived straightforwardly,
see below and (ii) for certain energy regimes it serves as a good approximation for the coupled $\ell$-wave CIR
\cite{giannakeas2012}.\par

Employing thus the physical $K$ matrix in the \lspwa{\ell}, where $\ell$ is not restricted to belong to a specific
symmetry class
%---
\begin{align}
  \Kp{oo,\ell}&=-i\alpha_\ell\Fo{\ell\ell}\quad,\textrm{with}
  \label{eq:kphys_single_pw}
  \\
  \alpha_{\ell}&=\frac{i\Delta_{\ell}}{1-i\Delta_{\ell}\Umat{\ell\ell}},
  \label{eq:alpha_ell}
\end{align}
%---
we calculate according to Eqs. \eqref{eq:transmission_fplus} and \eqref{eq:reflection_fplus}
the transmission- and reflection coefficients
%---
\begin{align}
  T^{(\ell)}_{\nnp}&=\delta_{\nnp}+\frac{\alpha_{\ell}^2\left(2\tr{\Fo{\ell\ell}}(\Fo{\ell\ell})_{\nnp}\delta_{\nnp}-(\Fo{\ell\ell})_{\nnp}^2\right)}{1-\alpha_{\ell}^2\tr{\Fo{\ell\ell}}^2}\\
  R^{(\ell)}_{\nnp}&=\frac{-\alpha_{\ell}^2(\Fo{\ell\ell})_{\nnp}^2}{1-\alpha_{\ell}^2\tr{\Fo{\ell\ell}}^2},
  \label{eq:tmatrix_closed_form}
\end{align}
%---
from which we derive in particular the total transmission coefficient $T^{(\ell)}_n=\sum_{\np<n_o}T^{(\ell)}_{\nnp}$,
when incident in channel $n$ which takes the appealing form
%---
\begin{equation}
  T_n^{(\ell)}=1-\frac{\tr{\Kp{oo,\ell}}(\Kp{oo,\ell})_{nn}}{1+\tr{\Kp{oo,\ell}}^2}
  \label{eq:total_transmission_n}
\end{equation}
%---
This quantity is in the focus of the following analysis since it encapsulate all the relevant scattering information for
both all the open and closed channels.\par

\subsubsection{General aspects of collisions in harmonic waveguides}
\label{sss:resonance_limits}
In this subsection we focus on the general behavior of the transmission coefficients for partial waves $\ell=0$ and
$\ell=1$ at total collision energies beyond the single mode regime. In particular we note, that the
separation of length scale induces an energy scale separation.
This means that even several quanta of the transversal excitation imply that the corresponding energy dependence of the scaled $s$-wave scattering length
is negligible. Due to the
increasingly narrow width of higher partial wave resonances, this simplification is not legitimate for $\ell\ge 1$.
In addition we should recall that all the scattering lengths used in the analysis below are analytically obtained via a Lennard-Jones 6-10 potential.\par
%---
\begin{figure}[h]
\includegraphics[width=0.45\textwidth]{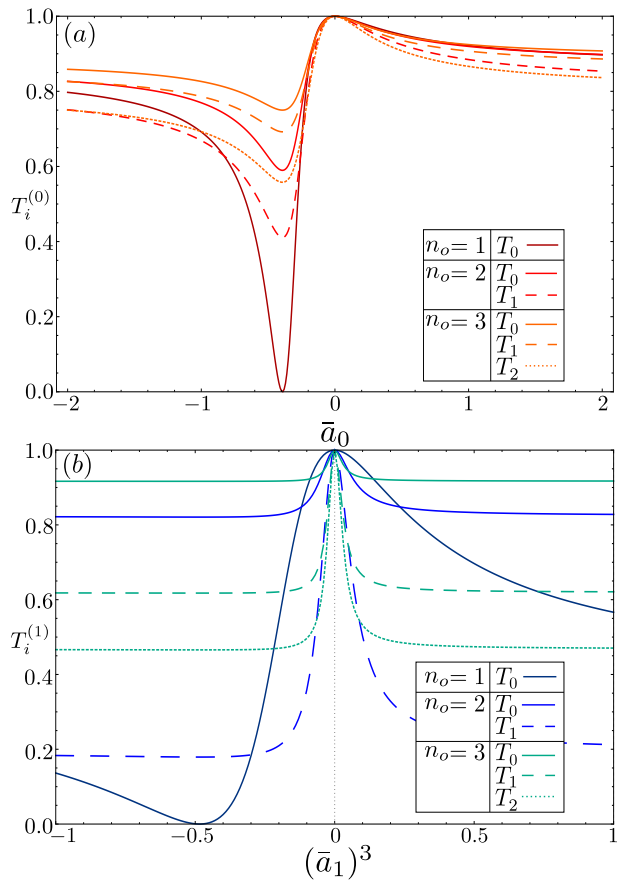}
\caption{(Color online) (a) Transmission coefficients $\ttc{0}{0},\ttc{0}{1},\ttc{0}{2}$ versus the
scaled $s$-wave scattering length $\abar{0}$, for the first, second and third open channel (solid, dashed and dotted
lines), respectively, at energy $\Delta\ep=0.95$. Higher lying curves for a particular number of open channels
correspond to a lower entrance channel. Panel (b) shows $\ttc{1}{0},\ttc{1}{1},\ttc{1}{2}$ at $\Delta\ep=0.05$ for
the case $\ell=1$.}
\label{fig:transmission_t123_boson_fermion}
\end{figure}
%---
First we show a typical case of the transmission coefficients $T_n^{(\ell)}$ for $\ell=0,1$ versus the
corresponding scaled scattering length $\bar{a}^{2\ell+1}_\ell$ in panels (a) and (b) of Fig.\ref{fig:transmission_t123_boson_fermion},
respectively. In particular we observe the asymmetric line shape of the transmission spectra for $\ell=0$. The
scaled energy above the corresponding channel threshold $\Delta\epsilon=0.95$ is chosen such that the CIRs, identified
as the minima of the respective transmission coefficients are best pronounced, i.e. we aim at a large difference
between the transmission values taken for large $|\bar{a}_0|$ and the specific value of $\bar{a}_0$ leading to a CIR.
This relative difference maximizes especially at energies below every threshold which can be readily seen in panel (a) of Fig. \ref{fig:transmission_bounds},
where the transmission coefficient values for $|\bar{a}_0|\to \infty$ ($T_{\infty,i}^{(0)}$ - black lines) and  for $\bar{a}_0$ at a CIR ($T_{\rm{CIR},i}^{(0)}$ - red lines) are displayed. Furthermore, we observe that a transmission
blockade is present only for the single mode regime (see Fig.\ref{fig:transmission_t123_boson_fermion} (a) red solid line). 
However, in the case of several open channels the transmission blockade is lifted giving in turn rise to
finite values of the transmission coefficient.
In addition transparency, i.e. $T_i^{(0)}=1$ occurs in the absence of interactions ($\bar{a}^{2\ell+1}_\ell=0$) between the colliding particles. Analogously, panel (b) of Fig. \ref{fig:transmission_t123_boson_fermion} depicts the results for $\ell=1$, as a function of the scaled $p$-wave scattering volume $(\bar{a}_1)^3$. Unlike the $s$-wave case, here the asymmetry of the line shape is barely visible if the
collision energy is raised above the first excited channel threshold.\par

%---
\begin{figure}[h]
\includegraphics[width=0.45\textwidth]{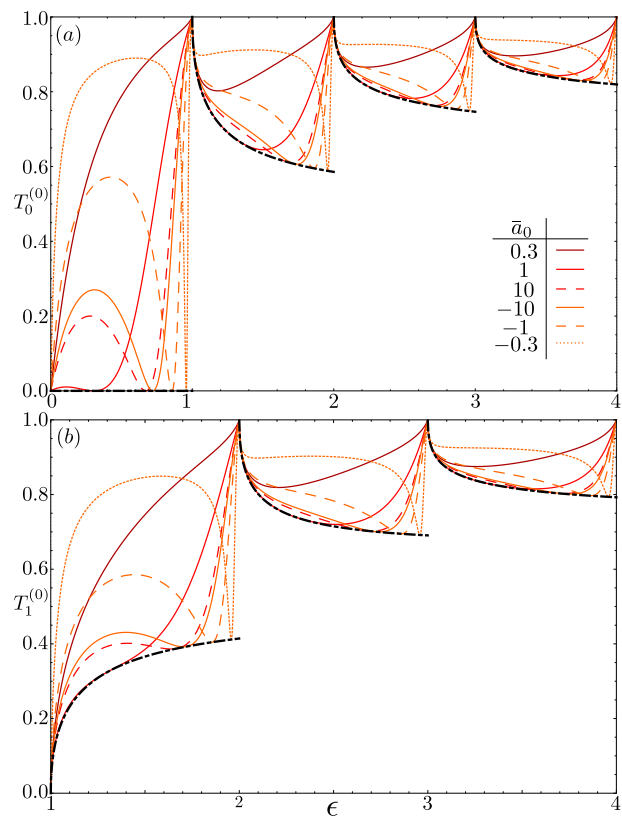}
\caption{(Color online) (a) Transmission coefficient $\ttc{0}{0}$ versus the channel scaled energy $\ep$
for $\ell=0$ for various scaled $s$-wave scattering lengths. The CIR obeys $\ttc{0}{0}=0$ in the ground
channel and for higher channels the resonances follow the confinement induced unitarity bound
$\tcir{0}{0}$ (thick, dot-dashed line), see Eq.  \eqref{eq:transmission_CIR}. (b) $\ttc{0}{1}$ versus $\ep$
for the same scattering lengths. Again, the resonances are bounded by $\tcir{1}{0}$.}
\label{fig:transmission_trend_s}
\end{figure}
%---
More insight into the behaviour of the transmission coefficients $T_i^{(0)}$ can be obtained from Fig. \ref{fig:transmission_trend_s},
where we present $T_0^{(0)}$ and $T_1^{(0)}$ versus $\epsilon$ in panels (a) and (b), respectively. These results
correspond to different values of the $C_{10}$ parameter. 
As it is discussed in more detail below (see Sec. IV A subsection 3 and 4), a CIR process belongs to
the transmission minima which occur in the interval of energies among the open channels.
Specifically, in Fig. \ref{fig:transmission_trend_s} (a) we observe that for $\epsilon<1$, ie between the thresholds of the ground and first excited transverse state, the transmission minima for varying values of the scattering lengths leads to a transmission blockade. For $1<\epsilon<2$
and higher energies the minimal value of the transmission is nonzero and increases with increasing channel the energy
belongs to. The transmission exhibits a repeating pattern shifted to higher values.
The locations of the minima with varying scattering length form thus a ``topos`` which in Fig. \ref{fig:transmission_trend_s} is denoted by the
black dot-dashed line: This is the {\it confinement-induced} (CI) unitarity bound and represents a universal feature.
We should note however that the specific functional form of the CI unitarity bound depends on the particular $\ell$-wave character of the collisions.
Furthermore, in Fig. \ref{fig:transmission_trend_s}(a) we observe that a CIR occurs for positive scattering length at $\Delta\epsilon < 0.69$, while CIRs of negative values of $\bar{a}_0$ emerge for $\Delta\epsilon > 0.69$.
The particular value of $\bar{a}_0=0.3$ does not permit a CIR at all and is hence monotonically increasing, while
departing from $\epsilon=0$ with a finite, non-vanishing slope. 

%---
\begin{figure}[h]
\includegraphics[width=0.45\textwidth]{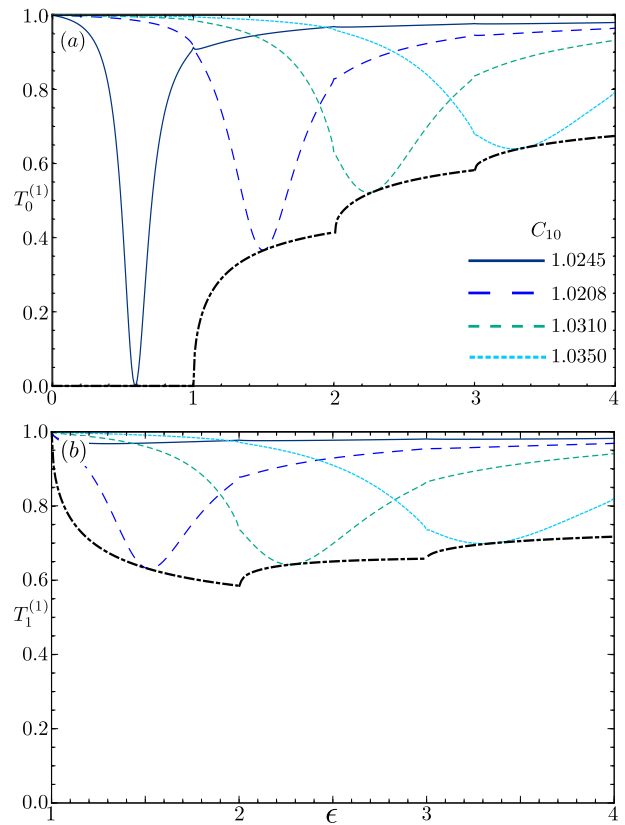}
\caption{(Color online) (a) $T_{0}^{(1)}$ versus the scaled energy $\ep$ for four open channels. Different curves
correspond to different values of the short range parameter $C_{10}$. It is clearly observed that the CIR saturates at
the confinement induced unitarity bound depicted as the dot-dashed curve. As for general $\ell$ a CIR associated with
a transmission blockade is only happening for $\ep\le 1$. In (b) the transmission coefficient $T_{1}^{(1)}$, when
incident in the first excited channel is shown.}
\label{fig:transmission_trend_p}
\end{figure}
%---
Finally we note that the transmission coefficient
goes up to unity at the channel thresholds regardless of the value of the scattering length. Considering the 
case $\ell=1$, presented in Fig. \ref{fig:transmission_trend_p}, we use the $C_{10}$
parameter to label the different curves. This is due to the narrow width of the $p$-wave free space resonances
which makes an energy independent treatment of the scaled $p$-wave scattering length impossible. For each curve the
$C_{10}$ parameter is adjusted such that a free-space resonance occurs for every energetic interval $i-1 \le\epsilon\le
i$ for $0\le i\le 4$. Again, at a CIR the corresponding transmission coefficient touches the $p$-wave CI unitarity
bound and thus we encounter a suppression of the transmission with a complete blockade $T=0$ for the case $\epsilon < 1$.
Away from a free-space resonance the scattering length quickly decreases to its small background value leading to
the large value of the transmission coefficient away from a resonance. Furthermore we observe that in the case of
$p$-wave interactions the value of $T_i^{(1)}$ taken at the channel thresholds indeed strongly depends on the scaled
$p$-wave scattering length, which drastically differs from the case of $\ell=0$ thus rendering the threshold
behavior for $\ell=0$ universal with respect to the $s$-wave scattering length.\par
\subsubsection{The decoupling energies}
\label{sss:decoupling_energy}
%---
\begin{figure}[h]
\includegraphics[width=0.45\textwidth]{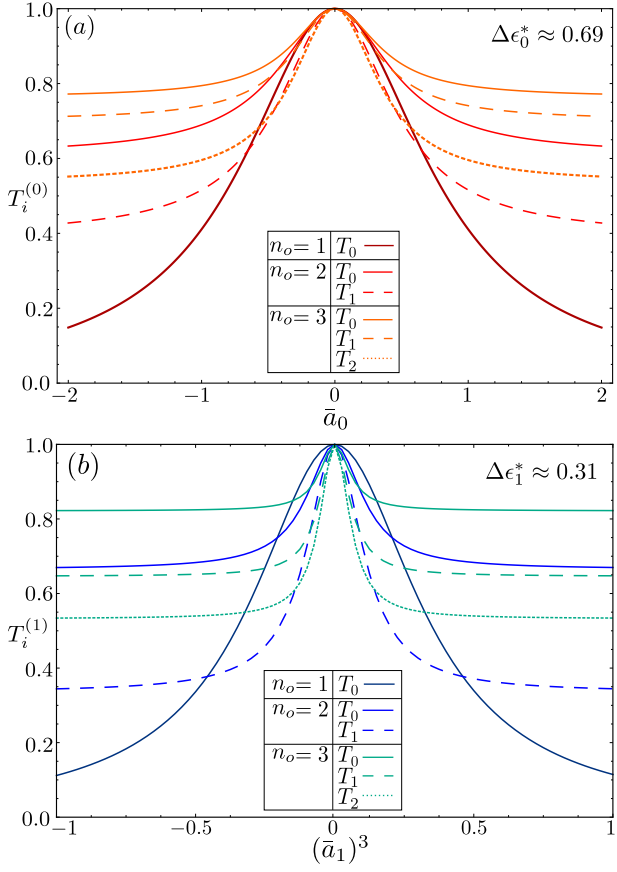}
\caption{(Color online) (a) Transmission coefficients at the decoupling energies $\lfloor
  \ep^*_0\rfloor+\Delta\ep^*_0$ versus the scaled $s$-wave scattering length $\abar{0}$. Different curves depict the
transmission coefficients for up to three open channels. The symmetric line shape centered around the non-interacting
case $\abar{0}=0$ indicates an effective decoupling from the closed channels of the waveguide. For large values of
$|\abar{0}|$, the transmission approaches the values given by Eq.  \eqref{eq:transmission_unitarity_temp}. The case of
$\ell=1$ is shown in (b) for energies $\lfloor \ep^*_1\rfloor+\Delta\ep^*_1$.}
\label{fig:transmission_decouple_sp}
\end{figure}
%---
In Ref. \cite{hess2014} it was shown that the coupling of the partial waves to the closed channels, given by the
$\Umat{\ell\ell}(\ep)$'s in Eq.\eqref{eq:closed_channel_umat} exhibits roots for all considered partial waves $\ell$.
These roots $\estar{\ell n}=\lfloor \estar{\ell n}\rfloor+\Delta\estar{\ell n}$, in the
following called \emph{decoupling energies}, depend on the partial wave $\ell$, and, in general, also on the number of
open channels $n$. The decoupling energies determine the particular energy where the bound state of the closed channels
decouples from \emph{all} the open channels. Therefore, in this case, the physical \kmat can be written in the
following form
%---
\begin{align}
  \Kp{oo}|_{\lfloor\ep\rfloor+\ep^*}&=(\KtD)_{\ell\ell}\;\Fo{\ell\ell}(\lfloor\ep\rfloor+\ep^{*})\nonumber\\
&=\mathbf{U}^T\KtD\mathbf{U}|_{\lfloor\ep\rfloor +\ep^*}
\label{eq:kmatrix_decoupled}
\end{align}
%---
where $\lfloor\ep\rfloor$ is the threshold energy for the largest open channel. Equation \eqref{eq:kmatrix_decoupled}
shows the expected result, that the 3D scattering information  which emerges close to the origin is transfered
to the asymptotic regime without being
affected by the closed channels of the trapping potential. In other words the colliding pair experiences effectively a
free space collision in the presence of the waveguide geometry.\par

In particular by using the expression for $\Umat{\ell\ell}(\ep)$ in Eq. \eqref{eq:closed_channel_umat} one can show,
that for the cases $\ell=0$ and $1$, the decoupling energies do not depend on the number of open channels, i.e.
$\Delta\ep^{*}_{\ell n}=\Delta\ep^*_\ell$, which means that the closed channels decouple from the open channels at the
same energy $\Delta\ep^*_\ell$ above the channel threshold, as it is illustrated in Fig.
\ref{fig:transmission_decouple_sp}. Fig. \ref{fig:transmission_decouple_sp}(a) shows the bosonic case, with
$\ell=0$ and $\Delta\ep^{*}_0\approx 0.69$, where the transmission coefficients $\ttc{0}{0},\ttc{0}{1},\ttc{0}{2}$
versus the scaled $s$-wave scattering length $\abar{0}$ are shown for one, two and three open channels, respectively.
The symmetric line shape centered around the non-interacting case $\abar{0}=0$ is clearly seen. Transmissions for
$\ell=1$ and $\ep^{*}_1\approx 0.31$, are shown in panel (b) as a function of the scaled $p$-wave scattering volume
$(\abar{1})^3$. In both panels the respective scaled scattering length are considered as external parameters which is in
the \lspwa{\ell} equivalent to a change of the $C_{10}$ parameter.

\subsubsection{The CIR limit}
\label{sss:cir_limit}

\begin{figure}[h]
\includegraphics[width=0.45\textwidth]{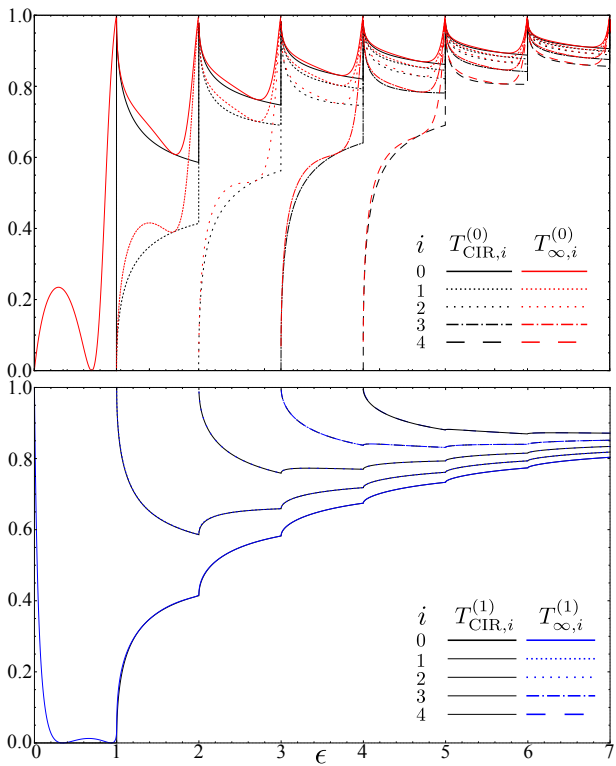}
\caption{(Color online) (a) Transmission coefficient $T_{\text{CIR},i}^{(\ell)}$ at a CIR, i.e. the
confinement induced unitarity bound, versus the
scaled energy $\ep$ as well as the transmission coefficient $T_{\infty,i}^{(\ell)}$ taken for large scattering length
$|\bar{a}_{\ell}|$, i.e. for scattering at the free-space unitarity bound. (b) Corresponding result for $\ell=1$.
Note that the two curves rapidly converge, implying a far less pronounced CIR in terms of
a transmission suppression than in the case of $\ell=0$.
}
\label{fig:transmission_bounds}
\end{figure}

As already emphasized above a CIR occurs in the vicinity
of a free-space resonance, i.e. in parameter regions $|\bar{a}_{\ell}|\gg1$. For this free-space unitarity regime the
scattering amplitude matrix in the \lspwa{\ell} is simply given by
%---
\begin{equation}
  \boldsymbol f_{\infty}^{(\ell)}=-\frac{\Fo{\ell\ell}}{\Umat{\ell\ell}+\tr{\Fo{\ell\ell}}},
  \label{eq:scattering_amplitude_unitarity}
\end{equation}
%---
from which we readily derive the corresponding transmission coefficients 
%---
\begin{equation}
  T_{\infty,n}^{(\ell)}=\frac{\Umat{\ell\ell}^2-\tr{\Fo{\ell\ell}}(\tr{\Fo{\ell\ell}}-(\Fo{\ell\ell})_{nn})}{\Umat{\ell\ell}^2-\tr{\Fo{\ell\ell}}^2}.
\label{eq:transmission_unitarity_temp}
\end{equation}
%---
This equation gives the values in the wings of large scattering length of the transmission
coefficients of Figs. \ref{fig:transmission_t123_boson_fermion} and \ref{fig:transmission_decouple_sp}. Furthermore,
its general energy dependence is show in panels (a) and (b) of Fig. \ref{fig:transmission_bounds} for the cases of
$\ell=0$ and $1$, respectively. There the transmission coefficients (ie. $T^{(\ell)}_{\infty,n}$) for large values of
the scattering length are compared
to the quasi 1D unitarity bound, i.e. the bound to the transmission coefficient at a CIR ie. $T^{(\ell)}_{\rm{CIR},n}$.\par

To understand the behavior of the transmission coefficient at a CIR, recall that the denominator of Eq.
\eqref{eq:kphys_single_pw} represents $\det(\openone-i\K{cc})$, which implies, that if this expression vanishes for a
particular value of $\Delta_\ell$, or the corresponding scattering length, respectively, a CIR occurs. We hence have a sufficient
criterion for the occurrence of a CIR given by the condition $\alpha_{\ell}\rightarrow\infty$, with $\alpha_{\ell}$
from Eq. \eqref{eq:alpha_ell}. A transmission value of
$T=0$ can only be achieved in the \lspwa{s} when there is a single open channel, since then
$\tr{\Fo{\ell\ell}}=\Fo{\ell\ell}$ holds. From a physical point of view this behavior is expected, since a
transmission blockade in a specific channel will not prevent the scattering into other channels, which
for example can be seen in Fig. \ref{fig:transmission_transistion_12_bosons}, where the transmission coefficients
$\ttc{0}{0}$ and $\ttc{0}{1}$ are shown, respectively. Both curves exhibit minima at the
CIR. Also in Fig. \ref{fig:transmission_transistion_12_bosons} we plot the transition coefficient $W_{01}^{(0)}$ (dotted
lines) which is enhanced at a CIR clearly demonstrating that for two open channels at CIR the inelastic process are
enhanced prohibiting in this manner the transmission coefficient to be zero when more than one channel is
involved.\par
%---
%---
The absence of a transmission blockade at a CIR in the case of more than one open channel, can be derived quantitatively
from the formal limit $\alpha_{\ell}\rightarrow\infty$ taken in Eq. \eqref{eq:connection_f_kphys}. 
This yields 
%---
\begin{equation}
  \boldsymbol f_{\textrm{CIR}}^{(\ell)}=-\frac{\Fo{\ell\ell}}{\tr{\Fo{\ell\ell}}},
  \label{eq:scattering_amplitude_CIR}
\end{equation}
%---
which generalizes the well known single mode result $f_{\textrm{CIR}}^{(\ell)}=-1$. It is remarkable
that for the transversal ground state a non-trivial energy dependence is not present. 
We note that Eq. \eqref{eq:scattering_amplitude_CIR} can, by virtue of \cite{hess2014}, 
also be obtained by formally equating Eq. \eqref{eq:scattering_amplitude_unitarity} at a vanishing closed channel coupling, i.e. $\boldsymbol
f_{\textrm{CIR}}^{(\ell)}=\boldsymbol f_{\infty}^{(\ell)}|_{\Umat{\ell\ell}=0}$. Writing down the transmission
coefficients for the scattering amplitude at a CIR, or, equivalently considering
$T_{\textrm{CIR},n}^{(\ell)}=\lim_{\alpha_{\ell}\rightarrow\infty}T^{(\ell)}_n$, yields
%---
\begin{equation}
  T_{\textrm{CIR},n}^{(\ell)}=1-\frac{(\Fo{\ell\ell})_{nn}}{\tr{\Fo{\ell\ell}}}.
\label{eq:transmission_CIR}
\end{equation}
%---
This expression in particular contains the previous statement that a CIR for a single open channel has
$T^{(\ell)}_0=0$.\par

In the \lspwa{\ell} the smallest value the transmission coefficient can take
as a function of the energy is determined by Eq. \eqref{eq:transmission_CIR}, which 
serves as a lower bound for the transmission coefficient.
This lower bound is compared to the value of the transmission coefficient for infinite
scattering length in Fig. \ref{fig:transmission_bounds}. 
The channel of incidence varies within the first few transversal modes, i.e. $0\le i \le 4$. As it
can be easily deduced from Eq. \eqref{eq:transmission_CIR} all quasi 1D unitarity bounds tend to unity irrespective of the
partial wave $\ell$ under consideration. This makes the CIR being less pronounced for increasing energies
(excited channels) with respect to the corresponding suppression of the
transmission. The difference between the transmission for infinite scattering length and at a CIR
is largest close to the channel thresholds for $\ell=0$. Remarkably, this difference 
$T_{\textrm{CIR},i}^{(1)}-T_{\infty,i}^{(1)}$ for $l=1$ happens to vanish as soon as the channel threshold to the first
excited mode is exceeded. This also explains the rather symmetric line shapes observed in panel (b) of Fig.
\ref{fig:transmission_t123_boson_fermion} for $n_o\ge 2$.\par

\begin{figure}[h]
\includegraphics[width=0.45\textwidth]{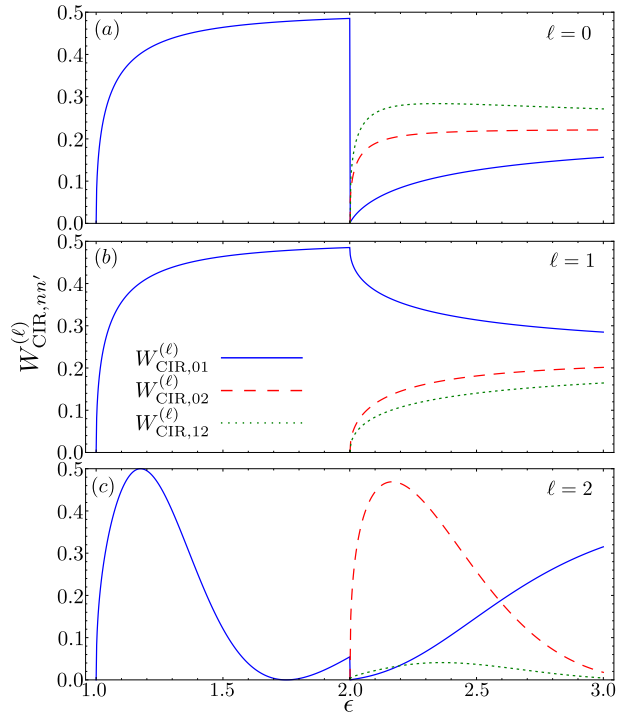}
\caption{(Color online) Confinement induced unitary bounds $W_{\textrm{CIR},nn^{\prime}}^{(\ell)}$ for
energies $\ep\le 3$. Panels (a)-(c) depict the cases $\ell=0,1$ and $2$, respectively. In particular,
$W_{\textrm{CIR},12}^{(\ell)}$ coincides for $\ell=1$ and $0$ for the first excited channel, as can be seen in panels
(a) and (b). The $d$-wave bound (c) exhibits a non-monotonic behavior as well as a total suppression of
transitions at $\ep=1.75$.
}
\label{fig:transitions_spd}
\end{figure}

A similar analysis can be performed in the \lspwa{\ell} for the transition coefficient $W_{\nnp}^{(\ell)}$. In this case, the
limit $\alpha_{\ell}\rightarrow\infty$ leads to
%---
\begin{equation}
  W_{\textrm{CIR},\nnp}^{(\ell)}=2\;\frac{(\Fo{\ell\ell})_{\nnp}^2}{\tr{\Fo{\ell\ell}}^2},
  \label{eq:transition_CIR}
\end{equation}
%---
describing the unitarity bounds for the transition coefficients which occur between different channels during a CIR.
Results for this observable are shown in Fig. \ref{fig:transitions_spd}, where all possible transition coefficients are
given for energies $\ep\le 3$. In particular, panel (a) and (b) depict the cases for $\ell=0$ and $1$, respectively.
In the energetic range $1\le\ep\le2$ the transition coefficients exactly coincide and exhibit a
monotonically increasing behavior. This coincidence abruptly changes when crossing the channel threshold to the
second open channel especially for $W_{\textrm{CIR},12}^{(0/1)}$, which has a sharp drop to zero for $\ell=0$ while for
$\ell=1$ the coefficient continuously decreases but remains the dominant transition process throughout the channel. 
Fig.\ref{fig:transitions_spd}(c),
depicts the case for $\ell=2$, which exhibits a non-monotonic behavior for all considered transition coefficients. Even
though, as discussed in more detail below in Sec. \ref{sss:impact_higher_partial_waves}, we do not expect this behavior
to exactly describe the processes for $d$-waves as additional $s$-wave contributions have to be taken into account, it is
nevertheless remarkable that for $\ell=2$ energetic regions exist where by means of a CIR no transitions between
channels can be induced, i.e. at the energy $\ep=1.75$ the transition probability between the two available channels
vanishes.\par

The interest in the observable $W$ is given by recent experiments on atom chips \cite{buecker2011,buecker2012,vanfrank2014}
where coherent excitations in higher transversal confinement modes are engineered. From this viewpoint the presented
analysis may contribute to an understanding in how far the CIR may be utilized to coherently excite atoms to higher
modes and which are the most efficient energetic regions in which these transitions can be achieved.\par

\subsubsection{Unitarity and CIR}
\label{sss:unitarity_and_cir}
To further illuminate the unitarity regime let us consider the situation from the viewpoint of
traditional scattering theory. Here, by using the appropriate relation for one spatial dimension
%---
\begin{equation}
  \mathbf{S}=1+2\boldsymbol{f}
  \label{eq:1d_rel_sMatrix_sAmplitude}
\end{equation}
%---
between the scattering matrix and amplitude \cite{lupusax1998thesis}, we readily derive the quasi 1D
\emph{unitarity relation}
%---
\begin{equation}
  \boldsymbol{f}\boldsymbol{f}^\dagger=-\;\mathfrak{Re}\,(\boldsymbol{f}),
  \label{eq:unitariry_relation}
\end{equation}
%---
where the right hand side denotes the real part of the scattering amplitude matrix, whereas the left hand side in
particular encapsulates the total reflection coefficient $R^{(\ell)}_n=\sum_{n\le n_o} R_{nn^{\prime}}^{(\ell)}$ with
$R_{n}^{(\ell)}=(\boldsymbol{f}\boldsymbol{f}^\dagger)_{nn}$. We thus conclude from the unitarity
relation, that the total reflection coefficient $R_n$ when incident in a specific channel $n$ is fully contained within
a single element of the scattering amplitude, namely
%---
\begin{equation}
  R_n^{(\ell)}=-\mathfrak{Re}\,(f_{nn}^{(\ell)})
  \label{eq:total_reflection_unitarity}
\end{equation}
%---
Equating now the right hand side with our system specific information from Eq. \eqref{eq:connection_f_kphys}, we find
after some algebra
%---
\begin{equation}
  R_n^{(\ell)}=\frac{(\Fo{\ell\ell})_{nn}}{\tr{\Fo{\ell\ell}}}\frac{1}{1+\frac{1}{(\alpha_{\ell}\tr{\Fo{\ell\ell}})^2}},
  \label{eq:reflection_n_unitarity}
\end{equation}
%---
which shows, that the scattering saturates at the unitarity bound $\alpha_\ell\rightarrow\infty$. It
is hence legitimate to regard $\tcir{n}{\ell}$ or the equivalent quantity $R_n^{(\ell)}|_{\alpha_{\ell}\rightarrow\infty}$ as the
\emph{confinement induced} (CI) unitarity bound, similar to the unitarity bound in free space, which scales as
$k^{-2}$.\par

\subsection{Transmission suppression in excited channels}
\label{sub:d-wave_blockade}

\begin{figure}[h]
\includegraphics[width=0.45\textwidth]{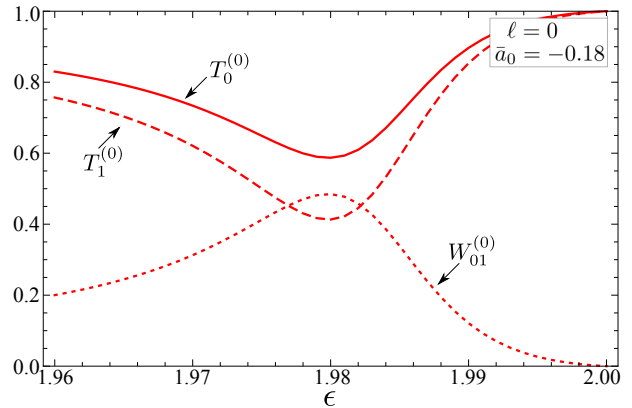}
\caption{(Color online) Transmission coefficient $\ttc{0}{0}$ (red solid line) for a small negative value of
the $s$-wave scattering length where the CIR is expected to occur near the threshold, namely at $\ep= 1.98$. At the same
position occurs also the CIR in the first excited channel see $T^{(0)}_1$ (dashed line).
The total blockade is absent, since the transition $W^{(0)}_{01}$ (dotted line) between the channels is also resonantly
enhanced at the position of the CIR.}
\label{fig:transmission_transistion_12_bosons}
\end{figure}
%---
\begin{figure}[h]
\includegraphics[width=0.4\textwidth]{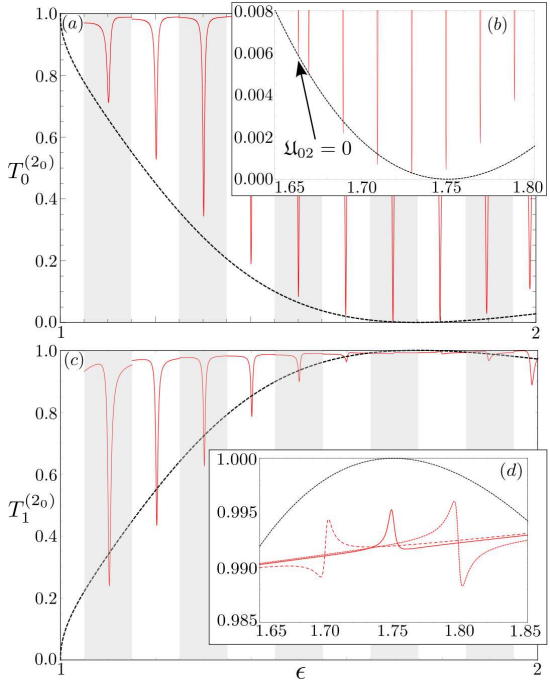}
\caption{(a) $T^{(2_0)}_0$ versus $\ep$ for coupled partial waves $\ell=0$ and $2$ constituted by different regions
of near resonant behaviour belonging to different $C_{10}$ values. In the middle of each vertical stripe a
free-space resonance translates to a corresponding CIR. The thick dashed line depicts $\tcir{0}{2}$. Deviations of the
transmission minima from the unitarity bound arise due to the \lspwa{d}.
The inset (b) shows a zoom-in plot around the blockade region ($T=0$) in the neighbourhood of $\ep\approx 1.75$. The leftmost
resonance (see arrow) coincides with the unitarity bound in which case the partial waves decouple, i.e. $\Umat{20}=0$. Panels (c)
and (d) show the results for $T^{(2_0)}_1$. Deviations from the \lspwa{d} become small around the decoupling
energy. We also observe that around this energy scattering becomes almost transparent.}
\label{fig:t1_dwave_comb_trend}
\end{figure}
%---
Increasing the total collision energy across the threshold of the first excited
transversal mode lifts in general the blockade in a
particular channel as the scattering constituents may escape via inelastic collisions to a different asymptotic state.
This is for example present in Fig. \ref{fig:transmission_transistion_12_bosons}, where a pronounced $s$-wave CIR is shown
for a total collision energy of $\ep\approx 1.98$. In particular, the transmission coefficients $T^{(0)}_0$ and
$T^{(0)}_1$ are shown together with the transition amplitude $W^{(0)}_{01}$ illustrating the resonantly enhanced
transition between the channels at a CIR.\par

Although Fig. \ref{fig:transmission_transistion_12_bosons} depicts the case of $s$-wave interactions it provides the
typical scenario when a CIR happens in the presence of more than one open channel. However, the first exception where
there is no resonantly enhanced transition between the channels at a CIR, occurs for $d$-waves when the particles are
incident in the first excited channel as we will demonstrate in Sec. \ref{sss:strong_surpression}. But before doing so,
we will provide a general brief discussion of the influence of higher partial waves.

\subsubsection{The impact of higher partial waves}
\label{sss:impact_higher_partial_waves}

The analysis of $s$- and $p$-waves scattering carried out in the \lspwa{\ell} was justified because of the negligible phase shifts
of allowed higher partial waves in the presence of a free-space resonance of $\ell=0$ and 1, respectively. This
particular simplification does of course depend on the considered interatomic potential, and in case of Lennard-Jones 6-10 potential this
certainly holds.
Nevertheless, when considering the CIRs associated to a higher partial wave, e.g. $d$-wave for the present discussion,
we have to take into account the non-vanishing $s$-wave scattering in the case of collisions between
indistinguishable bosons. This observation led to the idea of partial wave coupling due to closed channels of the
confinement, as firstly discussed in \cite{giannakeas2012}. The corresponding physical $K$ matrix of this problem based on
Eq. \eqref{eq:kphys} is now expanded in terms of $\Fo{\ell\ell}, \Fo{\elp\elp},\Fo{\ell\elp}$ and $\Fo{\elp\ell}$,
which serve as a basis for the two partial wave $K$ matrices. Therefore, it is not feasible to obtain a version of Eq.
\eqref{eq:tmatrix_closed_form} valid for two partial waves, since an inversion as needed in Eq.
\eqref{eq:connection_f_kphys} by means of a repeated application of the Sherman-Morrison method would yield a physical
\kmat with several hundreds of summands and is thus prohibitive. We therefore have to rely on a more qualitative analysis and take the
intuition from the transparent results obtained in the \lspwa{\ell}.\par

However, the coupling between the two contributing partial waves $\ell$ and $\elp$, mediated by the closed channels, is
conveniently described by the corresponding off-diagonal element $\Umat{\ell\elp}(\ep)$ of the closed channel coupling
matrix $\mathfrak{U}$. Similar to the previous case in Sec. \ref{sss:decoupling_energy}, where the vanishing diagonal
elements $\Umat{\ell\ell}$ were used to identify the decoupling energies for which the scattering process exhibits strong free-space character within the confinement, we generalize it here in the presence of a $\elp$-$\ell$ wave system. Our first observation is that there exist
also energies for which the off-diagonal elements of $\mathfrak{U}$ vanish, i.e. $\Umat{\ell\elp}(\ep_D)=0$ and the
confinement-induced coupling between the partial waves is absent. For the case of a coupled $s$-$d$-wave system, this
energy in the first excited channel is $\ep_D\approx 1.65$. In particular this also implies that the position of the
$d$-wave CIR is solely determined by the \lspwa{d}, i.e.
%---
\begin{equation} 
  \mathsf{RC}_{20}(\ep_D)=\mathsf{RC}_2(\ep_D),
  \label{eq:partial_wave_decoupling_}
\end{equation}
%---
where the resonance coefficients $\mathsf{RC}_{\ell}(\ep)$ and $\mathsf{RC}_{\ell\elp}(\ep)$ \cite{hess2014} are
described in Appendix \ref{app:notations}.\par

For the case of $s$- or $p$-waves, the coupling to the closed channels has a discrete shift
symmetry, i.e.
%---
\begin{equation}
  \Umat{\ell\ell}(\ep+N)=\Umat{\ell\ell}(\ep),\quad \mbox{with} \quad N\in \mathbb{N},
  \label{eq:translational_symmetry}
\end{equation}
which is equivalent to saying that the asymptotically defined transversal modes have no influence on the coupling
through the closed channels in the lowest possible partial wave for each respective class of exchange symmetry. The
origin of this symmetry is clearly the equidistant modes of the transversal harmonic oscillator in combination with the
particularly simple nodal structure of the local frame transformation for $s$- and $p$-wave interaction for the case 
$m=0$. This changes from the $d$-wave on. Since this feature does not depend on the coupling of partial waves, it can already be
understood in the \lspwa{d}, where a CIR occurs when $\abar{\ell}=\mathsf{RC}_\ell(\ep)$ is fulfilled. By inserting Eq.
\eqref{eq:closed_channel_umat} in Eq. \eqref{eq:rescon_single_appendix} one obtains
%---
\begin{equation}
  \frac{1}{\abar{\ell}^{2\ell+1}}=-i\;\sum_{p=0}^{2\ell}\bigl(2\sqrt{\ep+\tfrac{1}{2}}
  \bigr)^{2\ell-p}\cp{p}{\ell,\ell}\hurz{-\frac{p-1}{2}}{\epc},
  \label{eq:spwa_rescon_scatlength}
\end{equation}
%---
which depends on the total collision energy $\ep$ for partial waves other than $\ell=0$ and $\ell=1$,
since $\cp{0}{1,1}=\cp{2}{1,1}=0$. This rather technical observation of a channel dependent coupling to the closed
channels may indeed become relevant when experimentally trying to observe a $d$-wave CIR for collision energies $\ep>1$,
since the corresponding scattering length required to be comparable with the confinement length scale is reduced by the additional energy
dependent factor in Eq. \eqref{eq:spwa_rescon_scatlength}, which will likely make the $d$-wave CIR less difficult
to be observed at energies $\ep\gtrsim 1$. Having clarified this we can now focus our discussion of the transmission
suppression in the regime of multiple open channels.\par

\subsubsection{Strong suppression in the first excited channel}
\label{sss:strong_surpression}

Let us start the discussion on the strong transmission suppression in the first excited channel by a qualitative
analysis of the \lspwa{d} from which we will gain some insight. A sufficient and necessary condition for the occurrence
of a transmission blockade in the first excited channel ($1 \le \epsilon \le 2$) is of course a vanishing of the element $\tcir{n}{2}$, for
either $n=0$ or $1$, respectively. Indeed this happens in the \lspwa{d} for $\tcir{0}{2}$ at an energy $\ep_b=1.75$.
At the same energy $\tcir{1}{2}$ has to acquire an extremal value equal to unity. These two observations
imply that a CIR at $\ep_b$ will lead to a blockade for particles which are incident in the lowest transversal mode,
while for particles incident in the first excited channel, the CIR will result in complete transparency, i.e.
$\tcir{1}{2}=1$.\par

Taking additionally into account $s$-wave scattering which couples to the $d$-wave
this will influence the above-observed phenomena. The corresponding results are
presented in panels (a) and (c) of Fig. \ref{fig:t1_dwave_comb_trend}, where the two transmission
coefficients $\ttc{2_0}{0}$ and $\ttc{2_0}{1}$ are shown versus $\ep$ and the notation $2_0$ is used here to avoid confusion with
Eq. \eqref{eq:transmission_fplus} but to emphasize that the background contribution from the $s$-wave is properly taken
into account. Here, the individual resonances lying within the shaded vertical stripes result from changing the $C_{10}$
parameter. Due to the narrow width of the $d$-wave resonance only a relatively small window on the
energy axis is relevant to the collisional process since the scattering length rapidly decays back to a very small
value, resulting in a transmission close to unity. From these two panels one readily observes that the \lspwa{d} for the
CI unitarity bound $\tcir{0/1}{2}$, which is shown by the black dashed curve, is a rough approximation to the coupled
$d$-$s$-wave system.  However, by probing the CI unitarity bound for the coupled system we see that this approximation
qualitatively captures the observed values of the transmission coefficients at a CIR.\par

Furthermore, we observe in panels (b) and (d) of the same figure, that around $\ep_b$ the \lspwa{d} becomes more
reliable, which may be related to the decoupling of the partial waves, i.e. $\mathfrak{U}_{02}\approx 0$ also for
energies in that region. Nevertheless, the blockade expected in channel one from the \lspwa{d}, as well as the
transparency in channel two are not observed. This however is totally expected since the non-negligible $s$-wave
scattering length in the background prevents these extremal values. But still, for an extended region of energies
$1.6\le\ep\le 1.8$, the portion transmitted when incident in the ground channel is less than $1\%$. Similar, the transmitted
part when incident in the first excited channel is above $99\%$. Furthermore, the vanishing of the transmission
coefficient in the ground channel, i.e. $T_{0}^{(2_0)}=T_{00}^{(2_0)}+T_{01}^{(2_0)}\approx 0$ in particular also
implies that $T_{01}^{(2_0)}\approx 0$, which in turn leads to $W_{01}^{(2_0)}\approx 0$. This means that in this region
of energies the scattering processes preserve the channels, i.e. elastic collisions dominate since
the transition probability between the two open channels is negligible. We emphasize that this behavior corresponds to a
blockade for particles in the ground state while excited particles are effectively non-interacting regardless the fact
that there are two open channels. This observation might lead to interesting implications for the corresponding
many-body system, like a mixture of a Tonks-Girardeau gas and a non-interacting gas.

\subsection{Distinguishable particles}
\label{sub:distinguishable}
Let us start by briefly stating some single channel results, for which a similar analysis as in the case of indistinguishable
particles was carried out before. As introduced in Eq. \eqref{eq:transmission_fplus}, the appropriate scattering
observable for distinguishable particles (DP) is the transmission coefficient where both scattering amplitudes, one even
and one odd, are present.  The fact that the Hamiltonian of distinguishable particles still commutes with the
parity operator permits us to treat collisional events within the framework of the K-matrix approach. Therefore,
hereafter we will employ the Ansatz from Eq. \eqref{eq:transmission_fplus} with two single partial waves, namely $s$-
and $p$-wave, yielding
%---
\begin{align}
  T_{\nnp}^{(s,p)} &=\Bigl|\delta_{\nnp}+i\left(\Kp{oo,s}\left[\openone-i\Kp{oo,s}\right]^{-1}\right)_{\nnp}\nonumber\\
  &+i\left(\Kp{oo,p}\left[\openone-i\Kp{oo,p}\right]^{-1}\right)_{\nnp}\Bigr|^2,
\label{eq:transmission_distinguishable}
\end{align}
%---
$\Kp{oo,s}$ and $\Kp{oo,p}$ denote the corresponding $K$ matrices for the $s$- and $p$-wave, respectively. We note, that
likewise also both $K$ matrices with two contributing partial waves could be used, to describe a system where $s$-,
$p$-, $d$- and $f$-waves are significant. 
From the scalar, i.e. single channel version of Eq. \eqref{eq:transmission_distinguishable}, an effective physical \kmat
can be constructed which is explicitly given by
%---
\begin{equation}
K_{oo,\textrm{eff}}^{1D,phys}=\frac{K_{oo,s}^{1D,phys}-K_{oo,p}^{1D,phys}}{1+K_{oo,s}^{1D,phys}K_{oo,p}^{1D,phys}},
\label{eq:effective_k_matrix_distinguishable}
\end{equation}
%---
The construction is accomplished by simply demanding that the transmission coefficient takes the usual form of
$T=(1+K_{\textrm{eff}}^2)^{-1}$ \cite{giannakeas2012}. As in our previous study, this effective $K$ matrix is clearly separated in numerator
and denominator, which gives rise to transparency and resonance coefficients $\mathsf{TC}_{\ell\elp}$ and
$\mathsf{RC}_{\ell\elp}$, respectively. These coefficients are given by
%---
\begin{align}
  \mathsf{RC}_{\elp\ell}^{(\mathsf{dp})}(\ep)&=\frac{-1}{2\sqrt{\ep+\frac{1}{2}}}\sqrt[2\elp+1]{\frac{1}{i(\Umat{\elp\elp}+\alpha_{\ell}U_{\elp
0}^2U_{\ell 0}^2)}}
\label{eq:rescoeff_distin}
\\
\mathsf{TC}_{\elp\ell}^{(\mathsf{dp})}(\ep)&=\frac{-1}{2\sqrt{\ep+\frac{1}{2}}}\nonumber\\
  &\times\sqrt[2\elp+1]{\frac{\Delta_{\ell} U_{\ell 0}^2}{U_{\elp 0}^2-i\Delta_{\ell}(\Umat{\elp\elp}U_{\ell
  0}^2+\Umat{\ell\ell}U_{\elp 0}^2-2U_{\elp 0}^2U_{\ell 0}^2)}},
\label{eq:transcoeff_distin}
\end{align}
%---
where the superscript $(\mathsf{dp})$ abbreviates ``distinguishable particles'' to separate the notation from the one
used in Eqs. \eqref{eq:rescoeff_double_appendix} and \eqref{eq:transcoeff_double_appendix} for indistinguishable
particles. Here, $\ell$ and $\elp$ belong to different, namely even and odd, symmetries. The explicit form of these
coefficients is very similar to the ones introduced in \cite{hess2014} but, unlike them, do not contain off-diagonal
elements of the matrix $\mathfrak{U}$, which indicates the absence of couplings between even and odd partial waves due
to the confinement, becoming evident when comparing Eq.
\eqref{eq:rescoeff_distin} from above with Eq. \eqref{eq:rescoeff_double_appendix}, corresponding to indistinguishable
particles. If we interpret $U_{\elp 0}U_{\ell 0}$ as the trace over the open channels, which indeed can be shown
rigorously in a lengthy calculation for a problem involving multiple open channels, we see that the term proportional to
$\alpha_{\ell}$ in Eq. \eqref{eq:rescoeff_distin} is this trace squared, contrasting Eq.
\eqref{eq:rescoeff_double_appendix} where a analog squared trace over the closed channels is present. The same argument
also holds for the transparency coefficients Eqs. \eqref{eq:transcoeff_distin} and
\eqref{eq:transcoeff_double_appendix}, showing that the mechanism relating the partial waves is different for
indistinguishable and distinguishable particles, respectively. Whereas the former are coupled through the closed
channels, the latter are connected via the frame transformation in the open channels. Even though, there is an obvious structural
similarity between the corresponding resonance and transparency coefficients for distinguishable and indistinguishable
particles the physical mechanism coupling the partial waves is very different. Since the parity in $z$-direction is a
good quantum number the local frame transformation is not allowed to couple partial waves with
$\Delta\ell\pm 1$.  The appearing ``mixing term'' proportional to $\alpha_{\ell}$ in the coefficients
$\mathsf{TC}_{\elp\ell}^{(\textsf{dp})}$ and $\mathsf{RC}_{\elp\ell}^{(\textsf{dp})}$ originates thus from the coherent
superposition of the partial $s$- and $p$-wave contributions assumed in Eq.
\eqref{eq:transmission_distinguishable}.\par

%---
\begin{figure}[h]
\centering
\includegraphics[width=0.5\textwidth]{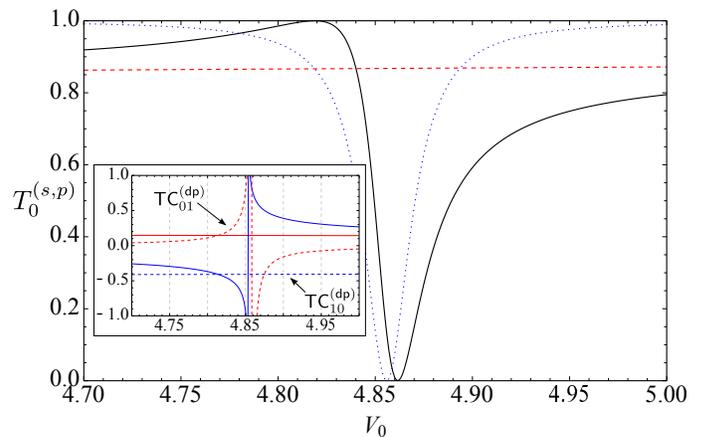}
\caption{Transmission coefficient $T_0^{(s,p)}$ versus potential depth $V_{0}$ of a spherical square
well. Black solid lines: with $s$- and $p$-wave interactions; red dashed line: $s$-wave approximation; blue dotted line:
$p$-wave interactions. The dual CIR is clearly seen to happen where the individual transmission coefficients of $s$- and $p$-wave interactions are
equal, as well as the shift of the $p$-wave CIR due to the non-negligible $s$-wave scattering. The inset shows the
scaled $s$- (red solid) and $p$-wave (blue solid) scattering lengths versus the potential depth. The crossings with the
DP transparency coefficients (dashed lines) determines the position of the dual
CIR.}
\label{fig:dual_cir_sp_zoom}
\end{figure}
%---
The arguments for the single channel case made above are illustrated in Fig. \ref{fig:dual_cir_sp_zoom}. Here, the black
solid line shows the transmission coefficient $T_0^{(s,p)}$ versus $V_{0}$, the depth of a spherical square well potential which
was used in order to model the two-body interactions of the DP. For comparison there are also the corresponding
transmission coefficients shown for the pure $s$- (red dashed) and $p$-wave (blue dotted) interaction, respectively. It
is clearly observed that a dual CIR corresponding to complete transparency \cite{kim2006,hess2014}
appears when the transmission coefficients of $s$ and $p$ are equal, i.e. by Eq.
\eqref{eq:effective_k_matrix_distinguishable} this means that the quasi 1D wavefunctions induced by s and p wave
interaction possess phase shifts of equal magnitude but differ by a sign yielding thus destructive interference.
As in \cite{kim2006}, this potential is used to mimic the possibility of having large values of $\abar{s}$ and
$\abar{p}$ simultaneously, illustrating the peculiar quasi 1D feature of total transparency while strongly interacting
with two partial waves. However, by inspecting the resonance and transparency coefficients $\mathsf{RC}^{(\mathsf{dp})}$ and
$\mathsf{TC}^{(\mathsf{dp})}$ we see, that it is not necessary to have large values for both scattering lengths, i.e. a
sufficient condition is one of them being large and the other one possessing a small but non-negligible value, as it was already
discussed in \cite{hess2014} for indistinguishable particles. In addition, the inset shows the intersection of the $s$-
and $p$-wave scattering length depicted by the red and blue solid lines with the corresponding transparency coefficients
$\mathsf{TC}^{(\mathsf{dp})}_{10}$ (red dashed line) and $\mathsf{TC}^{(\mathsf{dp})}_{01}$ (blue dashed line), respectively.\par

%---
\begin{figure}[h]
\includegraphics[width=0.45\textwidth]{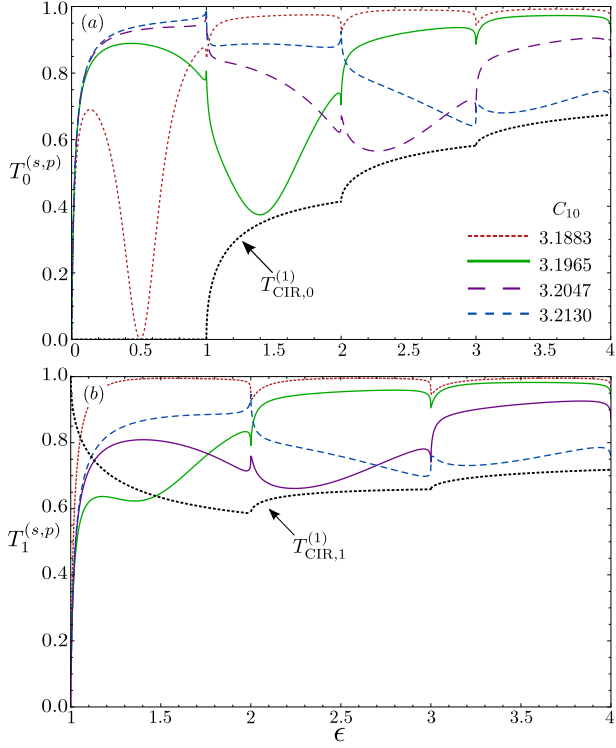}
\caption{ (a) Transmission coefficient $T^{(s,p)}_0$ versus the scaled energy $\ep$ (b) transmission coefficient $T^{(s,p)}_1$
when incident in the first excited channel. Similar to Fig. \ref{fig:transmission_trend_p},
individual curves correspond to different choices of the $C_{10}$ parameter in order to have resonanes in
every energetic interval $i-4\le\ep\le i$. In addition we show in both panels (a) and (b) the corresponding
CI unitarity bounds $T_{\textrm{CIR},n}^{(p)}$ (black dotted curve) for $n=0,1$.}
\label{fig:transmission_vs_energy_distinguishable}
\end{figure}
%---
Let us now address the multichannel scattering properties, based on Eq.\eqref{eq:transmission_distinguishable}.
We note, that an expanded representation for multiple open channels, similar to Eq.
\eqref{eq:tmatrix_closed_form}, in the single mode regime, is straightforwardly derived and thus not presented here. 
By considering a realistic short range potential as before, i.e. Lennard-Jones $6-10$ we have to distinguish two cases.
Firstly, the $s$-wave CIR in the presence of $p$-wave interactions and vice
versa.

An inspection of the corresponding scattering lengths in the former case shows that the $p$-wave scattering length in
the vicinity of a $s$-wave free space resonance is negligible, as it is the case for the indistinguishable $s-d$-coupled
case. Therefore, also for distinguishable particles,
$\mathsf{RC^{(dp)}_{01}(\ep)}\approx\mathsf{RC^{(dp)}_{0}}(\ep):=\mathsf{RC}_{0}(\ep)$, i.e.  the \lspwa{s} is a very
good approximation. Hence, the transmission coefficient versus energy would result in a behavior very much like the one
shown in Fig. \ref{fig:transmission_trend_s}. However, we note, that this simplification follows from our choice of
LJ-type potential, which, as already said before, is the adequate choice to describe the interactions of neutral atoms.
Nevertheless, in the vicinity of a free-space $p$-wave resonance the $s$-wave scattering length possesses a non-negligible
value and thus its contribution has to be taken into account.  The result on the transmission is shown in Fig.
\ref{fig:transmission_vs_energy_distinguishable}, where we consider the cases of incidence in the two lowest channels
for the case of four open channels. To obtain the different curves, the $C_{10}$ parameter is adjusted, such that there
is a free-space resonance of $p$-wave character at a particular energy. Focusing on the case of $p$-wave CIRs for
distinguishable particles, the black dotted line represents $\tcir{1}{1}$ in Fig. 10(a), while $\tcir{2}{1}$ is depicted
in Fig.10(b), showing that the neglect of a background $s$-wave scattering length leads to increasing deviations for
more open channels from the unitarity bound. However, we generally observe that the transmission is a mixture of $s$- and $p$-wave scattering.
For the $s-$wave part this is best seen by the zero energy value $T_1=0$ and the $\ep=1$ value of $T_{2}$ which is also zero. 
For the $p-$wave part we observe a transmission coefficient at threshold which is less than unity and
strongly depends on the $C_{10}$ parameter. These are two features we
find to be present in the \lspwa{p}. Therefore the contributions from both exchange
symmetries strongly contribute to the scattering physics of distinguishable particles.

\subsection{Threshold singularities}
\label{sub:threshold_singularities}

In Figs. \ref{fig:transmission_trend_s}, \ref{fig:transmission_trend_p} and
\ref{fig:transmission_vs_energy_distinguishable} we observe that at every channel threshold the transmission spectra
exhibit kinks. This constitutes another aspect of inelastic collisions, the so-called threshold singularities. More
specifically, it is known \cite{baranger1990,landau1981quantum} that when the total collision energy leads to
the opening of a new channel where new states become available this leads to a non analytic behavior of the
scattering matrix elements. \par

In order to firmly address this point it is useful to first consider the behavior of the individual elements
constituting the $K$ matrix when energetically approaching a channel threshold from below or above the closed channel
thresholds. Since the notion of transversal channels is inherently connected in our framework with the local frame
transformation $U_{ln}$ \cite{granger2004} and its derived quantity $\Umat{\ell\elp}$, the channel threshold behavior
will solely depend on these quantities and their properties around a corresponding threshold.  We start this analysis with the
local frame transformation $U_{ln}$, for which we find the following expression for the limit from below
%---
\begin{equation}
  \limU{\ep}{N}U_{lm}=\frac{(-1)^{d_0}\sqrt{2l+1}}{[4(N-m)(N+1/2)]^{1/4}}P_{l}(\sqrt{\frac{N-m}{N+1/2}}),
  \label{eq:limup_openchannel_U}
\end{equation}
%---
which holds for all open channels $m<N$, where $N$ denotes the threshold to the lowermost closed channel. From Eq.
\eqref{eq:limup_openchannel_U} we observe, that the elements of the local frame transformation acquire a finite value at
threshold and are continuous across the threshold. Next we inspect
the limit to the channel threshold from above for the element of the local frame transformation which
corresponds to the least open channel, i.e. $U_{\ell N}$, which will then become a closed channel when slightly further
decreasing the energy $\ep$. For this element we find the following expressions, which have to be distinguished for even
and odd partial waves
%---
\begin{align}
  \limD{\ep}{N}U_{\ell N}
&=C^{+}_{\ell N}\;\limD{\dep}{0}\Bigl(\frac{1}{\sqrt[4]{\dep}}+\bO{\dep^{\frac{3}{4}}}\Bigr),\label{eq:limdown_openchanel_U_even}\\
\limD{\ep}{N}U_{\ell N} &=C^{-}_{\ell N}\;\limD{\dep}{0}\Bigl(\sqrt[4]{\dep}+\bO{\dep^{\frac{5}{4}}}\Bigr),\quad
\textrm{with}
\label{eq:limdown_openchannel_U_odd}\\
C^{+}_{lN} &=\frac{\sqrt{2l+1}\;l!}{[(l/2)!]^2
  2^{l+\frac{1}{2}}(N+\frac{1}{2})^{1/4}}\nonumber\\
C^{-}_{lN} &=\frac{\sqrt{2l+1}\;(l+1)!}{(\frac{l+1}{2})!(\frac{l-1}{2})!
  2^{l+\frac{1}{2}}(N+\frac{1}{2})^{\frac{1}{4}}}\nonumber
\end{align}
%---
where Eqs. \eqref{eq:limdown_openchanel_U_even} and \eqref{eq:limdown_openchannel_U_odd} refer to the case of even and
odd partial waves, respectively. $C^+_{\ell N}$ and $C^-_{\ell N}$ denote constants depending on the number of open
channels as well as on the partial wave. Both equations exhibit a singular behavior at
threshold when approaching from above. In the case of even partial waves, this singularity is a pole, while
for the odd partial waves the local frame transformation becomes singular by means of an infinite slope at
$\epsilon=N$.\par

Next, let us consider the trace over the squared local frame transformations
$\mathfrak{U}_{ll^{\prime}}$ \cite{hess2014}. Here, we observe the following behavior for approaching the threshold from
above
%---
\begin{align}
\limD{\ep}{N}\Umat{ll^{\prime}}(\ep)
&=C^{(\ell,\elp)}_N+\limD{\dep}{0}\bO{\dep}\quad,\text{with}\label{eq:limdown_closedchannel_U}\\
C^{(\ell,\elp)}_{N} &=
\sum_{p=0}^{l+l^{\prime}}\frac{\cp{p}{\ell,\elp}}{(N-\frac{1}{2})^{\frac{p+1}{2}}}\;\zeta(-\frac{p-1}{2}),
\label{eq:limdown_closedchannel_U_coeff}
\end{align}
%---
where the coefficients $c_p^{(\ell,\elp)}$ are defined in Eq. \eqref{eq:clossel_channel_expansion_coefficients_cll}.
In Eq. \eqref{eq:limdown_closedchannel_U} we observe, that for the limit from above the elements of the matrix
$\Umat{}$ smoothly approach the channel threshold irrespective of the particle exchange symmetry. However, the actual
value attained does of course depend on the partial wave $\ell$, but there is no specific distinction between even and
odd partial waves. On the contrary the limit from below exhibits a more intricate behavior and is given by
%---
\begin{equation}
\limU{\ep}{N}\Umat{\ell\elp}(\ep)=C^{(\ell,\elp)}_{N}+\limD{\dep}{0}\Big(\sum_{p=0}^{\ell+\elp}\frac{\cp{p}{\ell,\elp}}{(N-\frac{1}{2})^{\frac{p+1}{2}}}\Bigr)(\dep)^{\frac{p-1}{2}},
\label{eq:limup_closedchannel_U}
\end{equation}
%---
where we observe that except from the constant value only terms contribute which have a fractional exponent in the
energy dependence, implying non-analytic behavior when approaching the channel thresholds from below.
We note that by comparing the local frame transformation $U_{\ell n}$ and the elements of the closed channel
coupling matrix $\Umat{}$, their limiting behavior is just reversed, i.e. one approaches the limit in a singular manner from above,
while regular from below and vice versa.\par

For the case of even partial waves, it can be shown by exploiting general
results on the Wigner $3j$-symbols that the coefficients $\cp{0}{\ell,\elp}$ are always
non-vanishing. This observation leads to the following asymptotic form
%---
\begin{equation}
\limU{\ep}{N}\Umat{\ell\elp}(\ep)=C^{(\ell,\elp)}_N+\frac{\cp{0}{\ell,\elp}}{\sqrt{N-\frac{1}{2}}}\limD{\dep}{0}(\dep)^{-\frac{1}{2}},
\label{eq:limup_closedchannel_U_bosonic}
\end{equation}
%---
which means that the $\Umat{\ell\elp}$'s diverge at threshold as $\frac{1}{\sqrt{\dep}}$.\par

Similar to the case of even partial waves we again exploit general properties of the Wigner $3j$-symbols to show for the
odd values of $\ell$ and $\elp$ that generally $c_{0}^{(\ell,\elp)}\equiv 0$, but $c_{1}^{(\ell,\elp)}\neq 0$.
This leads to the general from for the closed channel coupling, via
%---
\begin{equation}
\limU{\ep}{N}\Umat{\ell\elp}(\ep)=C^{(\ell,\elp)}_N+\frac{\cp{1}{\ell,\elp}}{(N-\frac{1}{2})}\limD{\dep}{0}(\dep)^{\frac{1}{2}}
\label{eq:limup_closedchannel_U_fermionic}
\end{equation}
%---
From Eqs. \eqref{eq:limdown_openchanel_U_even}-\eqref{eq:limup_closedchannel_U_fermionic} we observe
a fundamental difference between even and odd partial waves. While the elements $\Umat{\ell\elp}$
for even partial waves are discontinuous across a channel threshold, the odd counterpart is continuous.
However, we note that the limit from above in the odd case approaches the value
$C_{\ell\elp}$ with an infinite slope, which is equivalent to say, that $\Umat{\ell\elp}$ is not
Lipshitz continuous across threshold.\par

%---
\begin{figure}
\includegraphics[width=0.45\textwidth]{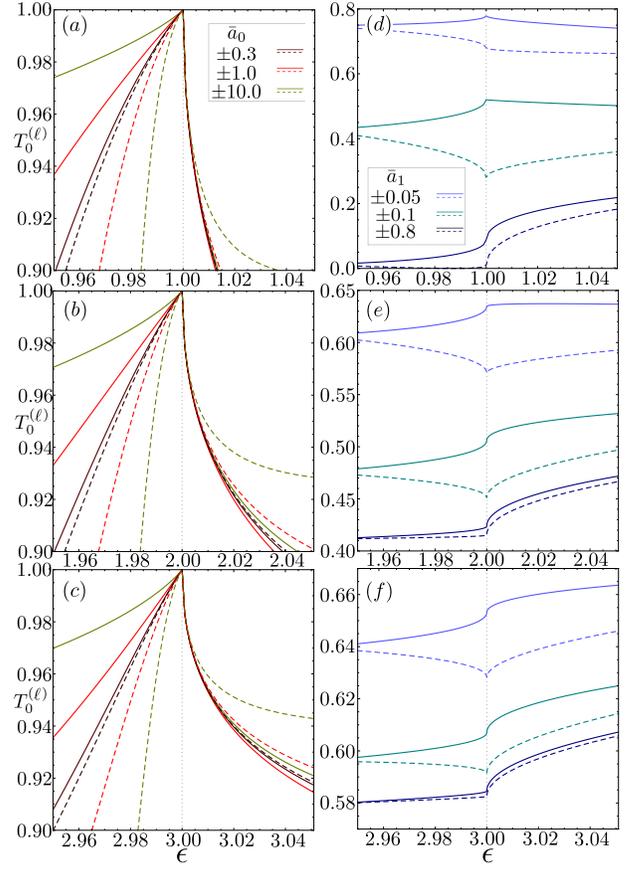}
\caption{(a) - (c) Transmission thresholds for the case $\ell=0$. Close to threshold the
  transmission approaches unity with a universal slope independent of the scattering length. This universal behavior is
  present for all even partial waves in the \lspwa{\ell}. (d) - (f) Transmission around thresholds is
  shown for a variety of $p$-wave scattering lengths, demonstrating the dependence on the scattering length around
  threshold.}
\label{fig:threshold_singularities_sp}
\end{figure}
%---
The different threshold behavior for indistinguishable bosons and fermions, i.e. even and odd partial waves, is clearly
observed in Fig. \ref{fig:threshold_singularities_sp}, where panels (a)-(c) depict the case for $\ell=0$ where, as
expected from the preceding analysis, the transmission obtains a universal slope and value around the thresholds,
regardless of the magnitude and the sign of the $s$-wave scattering length.  This is due to the divergent behavior of
$\Umat{\ell\ell}$ and $U_{\ell n}$ across threshold.  The trend exhibited in panels (a)-(c) is observed also for the
first few even partial waves in \lspwa{\ell}.  On the other hand, in panels (d)-(f) we illustrate the fermionic
counterpart for $\ell=1$. Here, we observe a richer structure of the threshold behavior due to the finite values the
quantities $\Umat{11}$ and $U_{1n}$ acquire. This circumstance renders the fermionic threshold behavior less universal
in the sense that the transmission value obtained does depend on the scattering length. On the other side, this
finiteness observed for odd partial waves allows for a CIR even at threshold. Notably the same behavior is apparent in
panels (d)-(f) for odd partial waves other than $p$-wave in the \lspwa{\ell} picture.

\section{Summary and Conclusions}
\label{sec:conclusions}
The focus of our present study is on the scattering behavior of identical bosons and fermions as well as on distinguishable
point-particles confined to a harmonic waveguide. The relevant coupling of different partial waves
as well as the explicit energy dependence of the scattering properties are properly taken into account.
The $K$ matrix formalism established in \cite{granger2004,giannakeas2012,giannakeas2013,zhang2013quasi1dscatt,hess2014} is employed to obtain the
relevant scattering observables. For higher partial wave interactions the explicit energy dependence of the scattering
lengths is properly taken into account by including the free-space scattering results of Gao
\cite{gao2009single,gao2000zero,gao1998qdtvdw}. Using these results we are able to present fully analytical results
including an adequate description for the interatomic scattering process.\par

We consider the scattering process under the assumption of scale separation of the length associated with the
interatomic and trapping potential, respectively. This assumption results in two regions of different symmetry, i.e.
spherical close to the origin and cylindrical in the asymptotic regime. The restriction that the total collision energy
lies below the threshold of the first excited transversal mode, as studied in previous works
\cite{olshanii1998,granger2004,giannakeas2012,hess2014} was dropped and we thus allow for inelastic scattering
describing transversal (de-)excitation processes. The relation between the quasi 1D scattering amplitude and the physical
$K$ matrix found in \cite{hess2014} was employed again to obtain analytical relations for the scattering observables.
This formalism allows for a unified treatment of inelastic collisions within a harmonic waveguide for distinguishable
and indistinguishable particles.\par

For the scattering of identical particles we investigated the transmission coefficient for up to four open channels
where we find good agreement in the single partial waves approximation (SPWA) with the numerical results derived earlier
\cite{saeidian2008}. For the \lspwa{\ell} we also derived a quasi 1D unitarity bound (CI unitarity bound)
which explains the influence of the open channels on the allowed transmission and transition coefficients around a CIR.
The universal aspects of the CI unitarity bound is demonstrated encapsulating all the relevant information of the interatomic interactions. 
However, the form of the CI unitarity bound depends of the exchange particle symmetry.
For the case of higher partial waves that are coupled we focus on the bosonic case where $\ell=0$ and $2$ waves
are coupled through the confinement. Here we studied the deviations from the \lspwa{d} to the case where the second
partial wave is taken into account. In particular we find that there is a region where the transmission when incident in
the ground channel almost vanishes, while when incident in the first excited channel the particles are non-interacting
to the same degree. We studied this behavior for the free-space phase shifts as energy dependent quantities by using the
analytic results of Gao \cite{gao2009single} on the scattering phase shift for potentials possessing a van der Waals
tail, relating to recent experimental observation \cs \cite{chin2004}, where $d$-wave shape-resonances were found.
Furthermore, similar to the decoupling from the closed channels, described before by a vanishing element
$\Umat{\ell\ell}$ we also find regions where the coupling between the partial waves vanishes, i.e. $\Umat{02}$ vanishes.
Over there, the \lspwa{d} is a good approximation to the transmission at a CIR. We note that also the corresponding
fermionic case can be treated within the same formalism. For the case of distinguishable particles we derive
resonance and transparency coefficients for the (dual) CIR. Here we observe
that the mechanism is way different from the case of indistinguishable particles where in particular the dual CIR was
achieved by interference of partial waves coupled through the closed channels, while in the distinguishable case the
coupling was accomplished by the open channels. Finally we provide a brief description of the origin and the type of the
appearing threshold singularities.

\begin{acknowledgments}
P.G. acknowledges financial support by the NSF through grant PHY-1306905. The authors thank C.H. Greene and V.
S. Melezhik for fruitful discussions.
\end{acknowledgments}

%-----APPENDIX-----
\appendix

\section{The physical $K$ matrix}
\label{app:kphys}
In order to analytically invert the contribution of the closed channels to the physical $K$ matrix,
$(\openone-iK^{1D}_{cc})^{-1}$, we employ the following method.
First, we recognize, that the $3D$ $K$ matrix, which is assumed to be diagonal, can be written as 
\begin{equation}
K^{3D}=\sum_{l=0}^{L}\Delta_{l}\mathbf{e}_{l}\mathbf{e}_{l}^{T},
\label{eq:3d_kmatrix_dyadic}
\end{equation}
here $\Delta_{i}=\tan\delta_{i}$ and $\mathbf{e}_{i}$
denotes the i-th Cartesian basis vector. Using now the relation $K^{1D}=U^{T}K^{3D}U$ between the
$K$ matrices in the different regions in configuration space, where $U$ denotes the local frame
transformation \cite{harmin1982nonhydrostark,*harmin1982stark,*harmin1985,fano1981,greene1987}, we obtain
%---
\begin{equation}
K^{1D}=\sum_{l}^{L}\Delta_{l}U^{T}\mathbf{e}_{l}\mathbf{e}_{l}^{T}U=\sum_{l=0}^L\Delta_l
\f{l}\f{l}^T,
\label{eq:1d_kmatrix_dyadic}
\end{equation}
%---
where we have introduced the frame transformed basis
%---
\begin{equation}
\mathbf{f}_i=U^{T}\mathbf{e}_i
\label{eq:lftransformed_basis}
\end{equation}
%---
For convenience we partition the frame transformed basis into two parts $\fo{i}$ and
$\fc{i}$, corresponding to open and closed channels, respectively and also introducing the
abbreviations $\Fo{ij}$ and $\Fc{ij}$ to denote the dyads $\fo{i}\otimes\fo{j}$ and
$\fc{i}\otimes\fc{j}$, respectively. We note, that this definition generalizes the open and closed
$K$ matrices attributed to a specific partial wave, since the diagonal elements of $\Fo{}$ and $\Fc{}$ are given by
%---
\begin{eqnarray}
\Delta_\ell\;\Fo{\ell\ell}&=& \K{oo,\ell}\\
\Delta_\ell\;\Fc{\ell\ell}&=& \K{cc,\ell}
\label{eq:generalization_k_to_dyad}
\end{eqnarray}
%---
In particular we encounter the following frequently appearing relation
%---
\begin{align}
  & (\K{oc,i}\Fc{jk}\K{co,l})_{n,n^\prime} = \nonumber \\ &= \sum_{\eta=n_c}^\infty\sum_{\eta^{\prime}=n_c}^\infty
  (\K{oc,i})_{n,\eta}(\Fc{jk})_{\eta,\eta^\prime}(\K{co,l})_{\eta^\prime,n^\prime}\nonumber\\
  &= \Delta_i\Delta_l\sum_{\eta=n_c}^\infty\sum_{\eta^{\prime}=n_c}^\infty
  U_{in}U_{i\eta}U_{j\eta}U_{k\eta^{\prime}}U_{l\eta^{\prime}}U_{ln^{\prime}}\nonumber\\
  &= \Delta_i\Delta_l
  U_{in}U_{ln^{\prime}}\Bigl(\sum_{\eta=n_c}^\infty U_{i\eta}U_{j\eta}\Bigr)\Bigl(\sum_{\eta=n_c}^\infty
  U_{k\eta}U_{l\eta}\Bigr)\nonumber\\
  &= \Delta_i\;\Delta_l\;\mathfrak{U}_{ij}\;\mathfrak{U}_{kl}\;(\Fo{il})_{n,n^{\prime}},
  \label{eq:KcoFccKco}
\end{align}
%---
where $n,n^{\prime}$ range within the open channels and the $\mathfrak{U}_{ij}$'s are the coupling elements derived
earlier \cite{hess2014}. These energy dependent elements are defined according to
\begin{equation}
  \Umat{\ell\elp}=\sum_{n=n_o}^\infty U_{\ell n}U_{\elp n},
  \label{eq:definition_umat_formal}
\end{equation}
which are given in Eqs. \eqref{eq:closed_channel_umat}, \eqref{eq:clossel_channel_expansion_coefficients_cll} and
\eqref{eq:clossel_channel_gamma} in closed form.
In a similar fashion to Eq. \eqref{eq:KcoFccKco}, the following relation is derived
%---
\begin{equation}
  \K{oc,i}\openone\K{co,j}=\Delta_i\Delta_j\;\mathfrak{U}_{ij}\Fo{ij}
  \label{eq:Koc1Kco}
\end{equation}
%---
Appearing higher order products of the closed channel $K$ matrices are readily shown to satisfy
%---
\begin{equation}
  \K{cc,i}\cdot\K{cc,j}\cdot\K{cc,k}=\Delta_i\Delta_j\Delta_k\;\mathfrak{U}_{ij}\mathfrak{U}_{jk}\Fc{ik}
  \label{eq:KcciKccjKcck}
\end{equation}
%---
The relations (A6) to (A8) turn out to be very useful when actually carrying out the analytical inversion of the matrix
$(\openone-i \K{cc})$. Similar to the procedure in \cite{giannakeas2012}, the inversion is done by
first recognizing, that the expansion of the 1D $K$ matrix given in Eq. \eqref{eq:1d_kmatrix_dyadic}
is written as a sum over dyads, i.e. rank one matrices and then repeatedly applying the
Sherman-Morrison formula \cite{press2007numrecipes}. The result of this procedure yields
%---
\begin{align}
(\openone-i\K{cc})^{-1} &=\openone+\alpha_i(\openone+\alpha_i\beta_{ji}\Umat{ij}^2)\Fc{ii}\nonumber\\
&+\beta_{ji}\Fc{jj}+\alpha_{i}\beta_{ji}\Umat{ij}(\Fc{ij}+\Fc{ji}),
\label{eq:inverse_closed_channel_contribution}
\end{align}
%---
where the coefficients $\alpha$ and $\beta$ are given by
%---
\begin{eqnarray}
\alpha_i&=& \frac{i\Delta_i}{1-i\Delta_i\Umat{ii}}
\label{eq:coupling_alpha_appendix}
\\
\beta_{ji}&=& \frac{i\Delta_j}{1-i\Delta_j(\Umat{jj}-\alpha_i\Umat{ij}^2)}
\label{eq:coupling_beta_appendix}
\end{eqnarray}
%---
These coefficients play a similar role as the couplings $g_1$ and $g_2$ defined in
\cite{giannakeas2012}. Inserting now the result of Eq.
\eqref{eq:inverse_closed_channel_contribution} into the equation for the physical $K$ matrix and
using the relation $\alpha_i\beta_{ji}=\alpha_j\beta_{ij}$ one ends up with
%---
\begin{align}
\Kp{oo}&=\frac{1}{\det(\openone-i
\KtD\Umat{})}\times\Bigl(\Delta_\ell\Fo{\ell\ell}+\Delta_{\ell^{\prime}}\Fo{\ell^{\prime}\ell^{\prime}}-\nonumber\\
&-i\Delta_{\ell}\Delta_{\ell^{\prime}}\bigl(\Umat{\ell^{\prime}\ell^{\prime}}\Fo{\ell\ell}+\Umat{\ell\ell}\Fo{\ell^{\prime}\ell^{\prime}}-\Umat{\ell\ell^{\prime}}(\Fo{\ell\ell^{\prime}}+\Fo{\ell^{\prime}\ell})\bigr)\Bigr),
\label{eq:kphys_appendix}
\end{align}
%---
representing the physical $K$ matrix interacting with two partial waves $\ell$ and $\ell^{\prime}$.
We note, that this formula only holds for indistinguishable particles, i.e. both partial waves have
to be either even or odd. In addition, we readily observe that the $K$ matrix given in Eq.
\eqref{eq:kphys_appendix} is real and symmetric as expected.

\section{The resonance and transparency coefficients}
\label{app:notations}
This section is considered as a brief summary of notions used here, which were introduced in
\cite{hess2014} in order to keep this presentation as self-contained as possible.\par
We remind that the formation of a CIR is described by a diverging physical $K$ matrix,
i.e. the roots of $\det(\openone-iK_{cc})$. As it can be seen clearly in Eq. \eqref{eq:kphys_appendix}
this divergence can only be achieved by a vanishing denominator. Equivalently, this can be expressed
by a divergence of the couplings, given in Eqs. \eqref{eq:coupling_alpha_appendix} and
\eqref{eq:coupling_beta_appendix}. Explicitly this means that a $\ell$-wave CIR occurs in the
SPWA when $\alpha_{\ell}$ diverges, and similar, a $\elp$-wave CIR occurs in the presence of
$\ell$-wave interactions, when $\beta_{\elp\ell}$ diverges. Parameterizing this divergence of
$\alpha_\ell$ in terms of the scattering length and energy, yields
\begin{align}
\abar{\ell}(\ep)&=\mathsf{RC}_{\ell}(\ep)
\label{eq:rescon_single_appendix}
\\
\mathsf{RC}_{\ell}(\ep)&=\frac{-1}{2\sqrt{\epsilon+1/2}}\times
\sqrt[2\ell+1]{\frac{1}{i
\mathfrak{U}_{\ell\ell}(\epsilon)}},
\label{eq:rescoeff_single_appendix}
\end{align}
and similar when the coupling between two partial waves has to be taken into account
\begin{align}
\abar{\elp}(\ep)&=\mathsf{RC}_{\elp\ell}(\ep)
\label{eq:rescon_double_appendix}
\\
\mathsf{RC}_{\elp\ell}(\ep)&=\frac{-1}{2\sqrt{\epsilon+1/2}}\times
\sqrt[2\ell^{\prime}+1]{\frac{1}{i\bigl(\mathfrak{U}_{\ell^{\prime}\ell^{\prime}}-\alpha_{\ell}\mathfrak{U}^{2}_{\ell\ell^{\prime}}\bigr)}}
\label{eq:rescoeff_double_appendix}
\end{align}

The so-called dual CIR, where the total transmission becomes unity, is obtained within the same framework by a vanishing
numerator of the physical $K$ matrix and is due to the matrix nature of $K$ in the case of
multiple open channels only expressible in terms of transparency coefficients
$\mathsf{TC}_{\elp\ell}$ only in the case of a single open channel. However, by analogous
arguments as for the resonance coefficients $\mathsf{RC}$, introduced above, the parametrization of
a dual CIR is then given by
%---
\begin{align}
\abar{\elp}(\ep)&=\mathsf{TC}_{\elp\ell}(\ep)
\label{eq:transcon_double_appendix}
\\
\mathsf{TC}_{\ell^{\prime},\ell}(\epsilon)&=\frac{1}{2\sqrt{\epsilon+1/2}}\nonumber\\
&\times \sqrt[2\ell^{\prime}+1]{ \frac{ \Delta_{\ell }U_{\ell 0}^2
}{U_{\ell^{\prime}0}^2-i\Delta_\ell(\mathfrak{U}_{\ell^{\prime}\ell^{\prime}}U_{\ell
0}^2+\mathfrak{U}_{\ell\ell}U_{\ell^{\prime}0}^2-2\mathfrak{U}_{\ell\ell^{\prime}}U_{\ell0}U_{\ell^{\prime}0})}},
\label{eq:transcoeff_double_appendix}
\end{align}
%---
where we note that due to the needed destructive interference, the dual CIR is only possible when
more than one partial waves are taken into account.

%------
\bibliography{litmc}
%------

\end{document}